\documentclass[twocolumn]{aastex61}
\usepackage[utf8]{inputenc}
\usepackage{graphicx}
\usepackage{amsmath}
\usepackage{natbib}

\usepackage{xcolor}
\newcommand{\mc}[3]{\multicolumn{#1}{#2}{#3}}

\received{Nov. 14, 2017}
\revised{Jan. 26, 2018}

\begin{document}

\title{The Dynamics of Tightly-packed Planetary Systems in the Presence of an Outer Planet: case studies using Kepler-11 and Kepler-90.}
\author[0000-0001-8214-5147]{A. P. Granados Contreras}
\affil{Department of Physics and Astronomy\\
The University of British Columbia\\
6224 Agricultural Rd.,\\
Vancouver, BC V6T 1Z1, CA}
\author[0000-0002-0574-4418]{A. C. Boley}
\affiliation{Department of Physics and Astronomy\\
The University of British Columbia\\
6224 Agricultural Rd.,\\
Vancouver, BC V6T 1Z1, CA}
\correspondingauthor{\'A. P. Granados Contreras}
\email{granados@phas.ubc.ca, acboley@phas.ubc.ca}

\shorttitle{Dynamics of STIPs with an Outer Perturber}
\shortauthors{{Granados Contreras} \& Boley}

\begin{abstract}
We explore the effects of an undetected outer giant planet on the dynamics, observability, and stability of Systems with Tightly-packed Inner Planets (STIPs). We use direct numerical simulations along with secular theory and synthetic secular frequency spectra to analyze how analogues of Kepler-11 and Kepler-90 behave in the presence of a nearly co-planar, Jupiter-like outer perturber with semi-major axes between 1 and 5.2 au. Most locations of the outer perturber do not affect the evolution of the inner planetary systems, apart from altering precession frequencies. However, there are locations at which an outer planet causes system instability due to, in part, secular eccentricity resonances. In Kepler-90, there is a range of orbital distances for which the outer perturber drives planets b and c, through secular interactions, onto orbits with inclinations that are $\sim16^\circ$ away from the rest of the planets. Kepler-90 is stable in this configuration. Such secular resonances can thus affect the observed multiplicity of transiting systems. We also compare the synthetic apsidal and nodal precession frequencies with the secular theory and find some misalignment between principal frequencies, indicative of strong interactions between the planets (consistent with the system showing TTVs). First-order libration angles are calculated to identify MMRs in the systems, for which two near-MMRs are shown in Kepler-90, with a 5:4 between b and c, as well as a 3:2 between g and h.
\end{abstract}


\section{Introduction}

The {\it Kepler} mission has demonstrated that planetary systems with multiple planets on short orbital periods are common. Roughly 
90\% of the Kepler candidates are found on orbits less than about 90 days \citep{Burke14}, with 23\% of Kepler stellar hosts 
containing more than one candidate. In terms of absolute distances and periods, these planetary configurations are much more compact 
than the Solar System's architecture, in which Mercury orbits the Sun in 88 days. Approximately 46\% of all candidates reside in 
known multi-planet systems \citep{Burke14}. Based on survey results, the frequency of planets at short orbital periods around 
solar-type stars is between about 30 and 50\% \citep{Howard12, Mayor11}, suggesting that these systems may be present around at least 
5\% of stars. While this may not represent the majority of planetary systems after a billion or more years of evolution, these Systems 
with Tightly-packed Inner Planets (STIPs) may represent a very common mode of planet formation, with instability and high collisional 
speeds potentially leading to system decay over long timescales \citep{Volk15}. During planet building, migration of outer planets 
could frustrate the formation of planets at short orbital periods \citep{Batygin15}, even if their formation would otherwise be 
common. Moreover, the presence of outer planets can affect the evolution of inner planetary systems, particularly if the outer system 
becomes unstable \citep{Mustill17}. Such a situation has been extensively studied in the context of the solar system's evolution \citep{Brasser09, Brasser13, Agnor12, Kaib16}.

STIPs offer a range of planetary configurations that can be used to test planet formation hypotheses. For example, at face value 
their low-eccentricity, closely-spaced configurations are consistent with disc migration. However, there is an under abundance of 
planets in or near commensurabilities, which is a fundamental prediction of convergent disc migration \citep[e.g.,][]{Hands14} without 
additional processes. Dynamical migration, such as planet-planet scattering followed by tidal evolution \citep[e.g.,][]{Rasio96, 
Ford08, Chatt08}, cannot obviously explain the known orbital architectures because planets in STIPs have low eccentricities and mutual 
inclinations \citep{Fabrycky14}. The situation becomes further complicated when considering star-planet spin-orbit alignments. Most 
STIPs with spin-orbit measurements show that the stellar spin axes are aligned with the normals to the planetary orbital planes 
\citep{Fabrycky09, TepCat}. Yet for several systems (e.g. HAT-P-07, KELT-17, Kepler-63, WASP-08, WASP-33), the stellar spin obliquity 
is large. Such misalignments could be caused by processes intrinsic to the star \citep{Rogers12,Rogers13} but it is not yet clear 
whether this can explain the observed population. Non-trivial disc-star or disc-environment interactions could instead force the 
planet-forming disc to be misaligned during planet building \citep[e.g.][]{Batygin12, Crida14, Lai14, Bate10}. The misalignment may 
also reflect pure dynamical interactions, in which a distant and highly misaligned companion causes an entire STIP to oscillate in 
inclination while keeping the mutual inclinations of the planets in the STIP small \citep{Kaib11, Batygin11, boue14a, boue14b}. In 
this sense, the STIP oscillates as a rigid system. The frequency of oscillation is a function of the total mass of the inner planets, 
the mass of the outermost body, and the separation of the perturber and the STIP. The Lidov-Kozai mechanism, which would normally 
cause coupled changes in a planet's orbital eccentricity and inclination, is quenched by planet-planet interactions \citep{Innanen97, Takeda08}. 

Finally, secular dynamics may also play a fundamental role in shaping the observed population of planetary systems, which is the 
subject of this paper. Secular eccentricity resonances can give rise to seemingly spontaneous system instability \citep{Hansen15, 
Volk15}, causing planet mergers and potentially erosion. Secular inclination resonances can place some planets on large inclinations, 
without the presence of a highly inclined perturber. This could cause the star to have a large obliquity relative to some planets in 
the system and could ensure that only a fraction of the system would be observed by transits directly. In this paper, we 
seek to address the following questions: Can planets in a STIP exhibit large mutual inclination variations while remaining otherwise 
stable? Can we find plausible examples of this mechanism in action? We also explore the overall stability of STIPs in the presence of 
a single outer planet for a range of orbital distances. 

This paper is organized as follows: in Section \ref{sec:stability}, we summarize general considerations for stability of planetary 
systems based on their multiplicity and planet separations. In Section \ref{sec:metho}, we describe the experimental overview and 
methodology used throughout this work. Section \ref{sec:results} provides the results from our simulations and secular analyses; we 
discuss our results in Section \ref{sec:discussion}, and finally, summarize our findings in Section \ref{sec:summary}.

\section{General Considerations for Stability}
\label{sec:stability}

A pair of planets in a two-planet system will always be stable if they are separated by $\Delta > 2.4 \left( {\mu_0+\mu_1} 
\right)^{1/3}$ \citep{Gladman93}, where $\mu_0$ and $\mu_1$ are the masses of the inner and outer planet relative to the star, 
respectively. The second planet's semi-major axis is thus $a_1 = a_0 (1 + \Delta )$, where $a_0$ is the inner planet's semi-major 
axis. Planet separations can also be described in units of mutual Hill radii, where the mutual Hill radius $R_{mH} = 0.5\, (a_0 + a_1) 
 \left[ (\mu_0+\mu_1)/3 \right]^{1/3}$. This gives $\eta \equiv (a_1 - a_0)/R_{mH}$, for which two-planet stability requires 
$\eta\gtrsim 3.46/(1+\Delta/2)$. For example, a two Jupiter-mass planet system around a solar-mass star must have planet 
separations of $3\:R_{mH}$ or greater.

Unfortunately, there is no planet separation limit to ensure stability at all times when the number of planets $N_P>2$. In these 
cases, we can only give a typical timescale for the planets to become orbit crossing \citep[e.g.,][]{Obertas17}. This timescale 
depends on the initial $\eta$ and the number of adjacent planets with that $\eta$ \citep[e.g.][]{Chatt08}. For Earth-mass planets with 
$N_P \gtrsim 3$, a mutual Hill radius separation of 10 can allow long-term stability \citep{Smith09}. Such spacings and planet 
multiplicities are relevant to STIPs: based on the confirmed systems in the \textit{NASA Exoplanet Archive}\footnote{Accessed on July 
20, 2017.}, there are 74 systems with a multiplicity $N_P > 3$, 26 with $N_P >4$, 6 with $N_P >5$, and 2 with $N_P >6$. Few STIPs have 
planets with $ \eta < 10 $, suggesting that the known planets should be stable over the lifetime of most stars 
\citep{Lissa14b,Obertas17}. However, secular interactions can lead to the disruption of a system, even if the planetary spacing alone 
suggests longterm metastability. A classic example is the evolution of Mercury in the Solar System, which has a small but 
non-negligible probability of being driven to orbit crossing with Venus \citep{Laskar94}.

Because many STIPs have high multiplicity (which we take to be $ N_P > 3 $ in this paper), we must ask whether a large fraction of the 
systems with only two or three planets are decay products themselves\footnote{In planet building through planetesimal accumulation 
and then giant impacts, this is always true to some extent. Here we are referring specifically to otherwise fully built planetary 
systems that achieve instability in less than approximate 1 Gyr.}. There are also observability considerations. In particular, the 
presence of outer perturbers can affect the observed planet multiplicity of transiting systems, which can have bearing on how we 
interpret planet-star misalignment \citep[e.g.][]{Winn05, Kaib11, boue14a, boue14b}, at least in part.

\section{Experiment Overview and Methodology}\label{sec:metho}

We investigate the observability and stability of STIPs by using direct numerical integration along with secular theory. First, we 
examine the behavior of Kepler 11 (K11) and Kepler 90 (K90, also known as KIC 11442793 and KOI-351) with and without an additional, 
perturbing Jupiter-like planet. When included, this perturber is placed exterior to the last known planet in each system, initially at 
5.2 au. Because the location of the Jupiter analogue affects the secular frequencies, we ``migrate'' the planet inwards to sample a 
broader range of frequency space. 

\begin{figure}[t]
	\centering
	\includegraphics[width=0.5\textwidth]{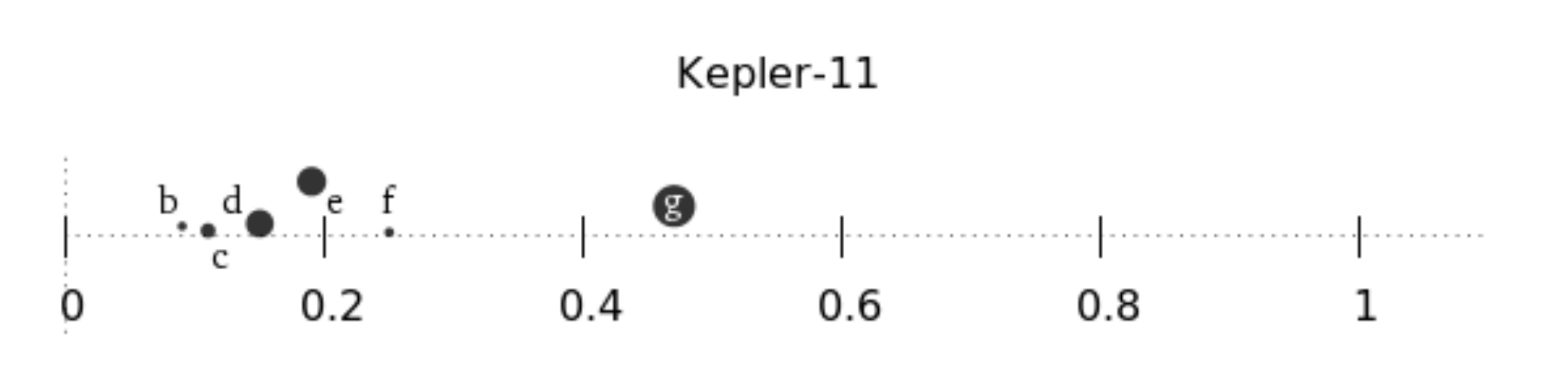}\\
	\includegraphics[width=0.5\textwidth]{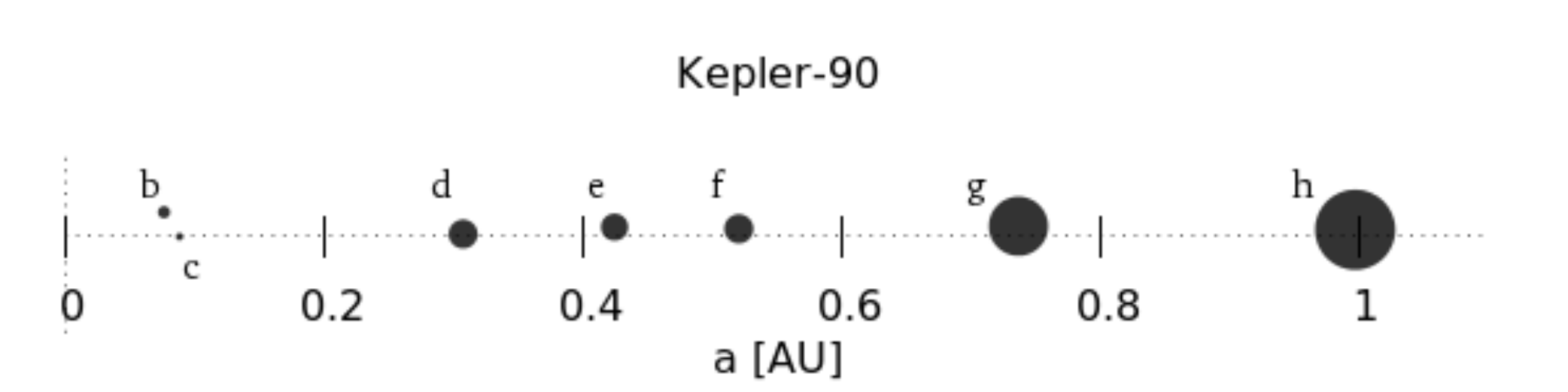}
	\caption{Spatial distribution of K11 and K90 systems, with the host star localized at the origin. In this plot, the size of each 
			planet represents its mass. In both cases, the mass of the planets seem to increase with their distance from the star. 
			The vertical offset represents the inclination of the planets.}
	\label{fig:0}
\end{figure}

K11 \citep{Lissa11,Lissa13} is composed of 6 planets that have orbital semi-major axes less than 0.5 au. The mass and orbital 
parameters of K11 planets are well defined, with the exception of the outermost planet, K11g, for which the mass and eccentricity only 
have upper limits, constrained by stability analyses \citep{Lissa13}. The estimated age of K11 is $9.7 \,\pm\, 1.5$ Gyrs, and 
the star has a mass of $M_* = 0.975 \,\pm\, 0.031\, M_\odot$ and a radius of $R_* = 1.193 \,\pm\, 0.115\, R_{\odot}$ \citep{Lissa13}. 
The other system, K90 \citep{Lissa14,Cab14,Schmitt14}, has 7 planets orbiting an $M_* = 1.2 \,\pm \,0.1 \,M_\odot$ star\footnote{In Dec.~2017, after the submission of this paper, an 8th planet candidate in K90 was announced \citep{Shallue17}. We are mainly using K11 and K90 as templates for STIPs, so the additional planet candidate does not directly affect the results presented here.}. The K90 
system (along with TRAPPIST-1) has the largest number of confirmed planets thus far, making it one of the closest in planet 
multiplicity to the Solar System. However, unlike the Solar System, all of K90's known planets are confined inside 1 au. The masses of 
K90's planets have not been measured directly, and as a result, the masses used here are estimated from the mass-radius relation 
described by \citet{exoorg1}, which is based on the size distribution reported by \citet{Lissa11b}. Table \ref{tab:1} summarizes the 
measured or estimated properties for K11 and K90 and Figure \ref{fig:0} emphasizes their relative spacing, mass, and inclination. The values in Table \ref{tab:1} are adopted for the present study, unless otherwise noted.

\begin{deluxetable}{cccccc}
\tablecaption{Nominal orbital elements of the known planets of K11 and K90. \label{tab:1}}
\tablehead{
\colhead{Planet} & \colhead{Mass} & \colhead{$a$} & \colhead{$e$} & \colhead{$i$} & \colhead{$\omega$} \\
\colhead{} &   \colhead{($M_\oplus$)}   & \colhead{($au$)} &   \colhead{}    & \colhead{($^\circ$)} &   \colhead{($^\circ$)}
}
\startdata
Kepler-11 & \mc{2}{c}{$M_* = 0.961 \,M_\odot$} & \mc{2}{c}{$R_* = 1.065 \,R_\odot$}  \\ \hline
		    b     &                1.9                 &                0.09                & 0.045 & 0.12 &   45.0   \\
		    c     &                2.9                 &                0.11                & 0.026 & 0.07 &   51.3   \\
		    d     &                7.3                 &                0.15                & 0.004 & 0.15 &  146.3   \\
		    e     &                8.0                 &                0.19                & 0.012 & 0.63 &   90.0   \\
		    f     &                2.0                 &                0.25                & 0.013 & 0.05 &   90.0   \\
		    g     &                20.0                &                0.47                & 0.100 & 0.35 &   90.0   \\ \hline
		Kepler-90 &  \mc{2}{c}{$M_* = 1.2 \,M_\odot$}  &  \mc{2}{c}{$R_* = 1.2 \,R_\odot$}   \\ \hline
		    b     &                2.4                 &               0.076                &   ...   & 0.28 &    ...     \\
		    c     &                1.7                 &               0.088                &   ...   & 0.00 &    ...     \\
		    d     &                7.9                 &               0.307                &   ...   & 0.03 &    ...     \\
		    e     &                6.9                 &               0.424                &   ...   & 0.11 &    ...     \\
		    f     &                8.1                 &               0.520                &   ...   & 0.09 &    ...     \\
		    g     &                69.1                &               0.736                &   ...   & 0.12 &    ...     \\
		    h     &               297.9                &               0.996                &   ...   & 0.08 &    ...     \\
\enddata
\tablecomments{The stability of Kepler-11 is very sensitive to the values of the argument of pericenter for planets e, f, and g.}
\end{deluxetable}

Published eccentricities and inclinations are used when available; otherwise, the orbital eccentricities are set to zero ($e=0$), 
particularly for K90. The inclination values given in Table \ref{tab:1} are relative to a reference plane. The plane was determined by 
averaging the published orbital inclinations relative to a perpendicular to the line of sight to the system, which is 
$<i_{\mathrm{K11}}> = 89.52^\circ$ and $<i_{\mathrm{K90}}> =89.68^\circ$. We create 100 realizations of each system, in which the 
longitudes and anomalies ($\Omega$, $\varpi$ and $\mathcal{M}$) for each planet were drawn from a uniform random distribution between 
0 and $2\pi$. These realizations are first run without the external perturber to establish the stability of the systems over 10 Myr. 
As will be described in the results (Section \ref{sec:results}), the stability of K11 is very sensitive to changes in the argument of 
pericenter. To further test the stability of K11 and K90 in the presence of a Jupiter analogue, we run the same initial conditions 
with an additional planet, also for 10 Myrs. Unless otherwise noted, the perturber has a mass $M_P=1 \,M_J$ and initial orbital 
elements $a_P = 5.2$ au, $e_P=0.05$, $i_P=1.3^\circ$, $\omega_P=273.8^\circ$, $\Omega_P=100.5^\circ$ and $\mathcal{M}_P=93.8^\circ$, 
where the subindex $P$ denotes parameters for the perturber. The realizations that include the perturber will be denoted as K11+ and 
K90+ throughout this work. The stability of the STIP is expected to be sensitive to the semi-major axis of the perturber (due to the 
secular frequencies), and as such, we systematically explore the semi-major axis parameter space of the perturber by forcing it to 
migrate inwards. The simulations that include the inward migration of the perturber are performed using the initial conditions that 
resulted in stable configurations for K11+ and K90+ after 10 Myrs. For K11+ we used the full set of stable realizations, but 
chose only one of the K90+ stable systems. This choice was based on the analysis of the K11+ simulations, which all showed consistent behavior.

N-body integrations were run using a modified version of the Mercury6 code \citep{mer6}, which includes a general relativity 
correction due to the proximity of the planets to their host star. The correction used here takes the form
\begin{align}
	{\bf a}_{\rm GR} & = - \frac{6 \,G^2 M^2}{r^4 c^2} \:{\bf r},
\end{align}
which is appropriate for low-eccentricity orbits \citep{GR1}. Because we want to resolve close encounters should they occur, we use 
the hybrid integrator of Mercury6. When in MVS mode, the time step is set to $10^{-2}$ of the initial period of the innermost planet. Again, the total integration time, if not otherwise stated, is 10 Myrs.
 
The inward migration of the Jupiter analogue is achieved by applying an acceleration term to the planet of the form
\begin{align}
{\bf a}_{\rm mig} & = -\frac{2\pi}{ \tau_{\rm mig}} \left( \frac{ 1 {\rm au}}{a} \right) \left[ 3 ( \bf{\hat{r}} \cdot {\bf \dot{r}} 
) 
{\bf 
\hat{r}} + ({\bf \hat{r}} \times {\bf \dot{r}}) \times{\bf \hat{r}}\right] \rm\nonumber \\
& = -\frac{2\pi}{ \tau_{\rm mig}} \left( \frac{ 1 {\rm au}}{a} \right) \left[ 2 \left( {\bf \hat{r} } \cdot {\bf \dot{r}} \right) 
{\bf 
\hat{r}} + 
{\bf \dot{r}} \right],
\end{align}
where $\bf r$ is the perturber's position vector from the star, $\bf \dot{r}$ its velocity vector, $a$ its semi-major axis, and 
$\tau_{\rm mig}$ is the timescale for migration at 1 au. Over the region of interest, migration is nearly constant. For these 
simulations, we set $\tau_{\rm mig}= 30$ Myr which is equivalent to $\dot{a} \approx 0.42$ au/Myr. During the 10 Myr migration, the orbital precession of the perturber increases by 1 to 2 orders of magnitude. The perturber's eccentricity also changes, decreasing among the stable systems from 0.05 to $\sim 0.035$ at $a_J=3$ au and to $0.02$ at $a_J = 1$ au, corresponding to $t = 6$ and 10 Myrs, respectively.
 
We complement the numerical simulations with secular theory as laid out by \citet{Murray99}. The theory considers only the secular 
contributions of the disturbing function to the equations of motion, expanded to second order in eccentricity, $e$, and inclination, 
$i$, as well as to first order in planetary mass. For systems in which the planets have small $e$ and $i$, the theory can identify 
locations at which a test particle would be in an eccentricity or inclination secular resonance, although higher-order methods 
\citep[e.g.,][]{Laskar85, Laskar86} or direct N-body calculations are needed to determine the outcome of a given resonance. If planets 
have strong interactions, such as what might be expected near mean motion resonances, then additional frequencies can become important 
and/or the predicted frequencies can become shifted with respect to the second-order theory. Because K11 and K90 exhibit TTVs, we do 
expect deviations from second-order theory, as is the case in the Solar System with Jupiter and Saturn \citep{Brouwer50}. 
Nevertheless, such effects are excluded in our calculations. We also exclude the effects of GR on the secular frequencies. Regardless, 
as will be shown in the results and discussion sections, secular theory seems to identify the locations of resonances with reasonable accuracy.

After building the second-order secular theory for any given planetary system configuration (e.g., the current STIP and perturber's 
orbital elements), we use the theory to highlight the resonant structure in the system. First, we introduce a test particle over a 
range of semi-major axes to sample the forced eccentricities and inclinations. Resonances will correspond to distances at which the 
forced eccentricity or inclination show very large increases, resulting from a singularity introduced when a precession frequency 
matches a system eigenfrequency. Using this approach, we can compare the resonant structure from secular theory with the behavior of 
the N-body simulations. For example, if a given planet becomes orbit crossing or develops large inclination variations, we can build 
the secular theory for the system, \textit{but exclude the highly perturbed planet(s)}. In K90(+), the two innermost planets are the 
most easily excited, so we examine the resonance structure by removing these planets from the secular theory, allowing us to see if 
the forced eccentricity and inclination at their semi-major axes suggest a resonance, at least for a test particle. This is only 
reasonable whenever the removed planets have low masses compared with the other planets. K90b and K90c are tightly coupled, so this 
must be done with some caution. Nevertheless, as we will show, this approach is reasonable enough to be useful for the present 
situation. As with K90(+), we examine the secular resonant structure in K11(+) by removing K11b, which is often the first planet to 
become orbit crossing, or by removing K11b and K11c as the planets are coupled. The eigenfrequencies for the second-order theory are 
obtained using a version of the secular theory python script available at \url{https://github.com/norabolig/resmap}.

We also analyze the resonant structure of K11(+) and K90(+) and examine the effects of strong planet interactions on the secular 
theory by calculating the synthetic frequency spectra for select systems. This is done by rerunning the given simulation for 0.3 Myrs with output every 60 years. The synthetic spectra are found by taking the discrete Fourier transform of 
\begin{gather}
e \, \exp(j \varpi) \label{eq:apsidal} \\
\sin (i) \,\exp(j \Omega)  \label{eq:nodal},
\end{gather}
for each planet, where $e$ is the eccentricity, $i$ the inclination, while $\varpi$ and $\Omega$ are the longitudes of pericenter and 
ascending node, respectively, and $j = \sqrt{-1}$. The Fourier transform of Eq. \ref{eq:apsidal} and Eq. \ref{eq:nodal} return the 
principal apsidal and nodal precession frequencies and amplitudes that compose the synthetic secular theory.\\
\section{Results} 
\label{sec:results}

The N-body simulations of K11 and K90 (without the perturber) allow us to examine the dynamical stability of our realizations of these 
systems. We find that the stability of K11 is very sensitive to perturbations to the nominal argument of pericenter whenever we 
consider non-zero eccentricity. In particular, 44\% of the simulations become unstable when using the nominal eccentricities but 
randomizing the longitudes. This is qualitatively consistent with \citet{Maha14}, who found that K11 is preferentially unstable if any 
of the eccentricities $e > 0.04$, using the same planetary masses used here. In contrast, 100\% of the realizations in this study are 
stable if the initial eccentricity is zero, which agrees with the zero-eccentricity stability tests of \citet{Lissa11,Lissa13}. 
For K90, 20\% of the our realizations present instability despite having initial orbital eccentricities set to zero for all planets. 
This result is unexpected at face value, as we might anticipate the system to be stable (as with K11) under such conditions. This 
agrees with the stability test of K90's discovery paper \citep{Cab14}. The system's secular perturbations could be the origin of the 
noted instability and/or mean motion resonances, as will be discussed later.
\begin{figure}[t!]
	\centering
 	\includegraphics[width=0.47\textwidth]{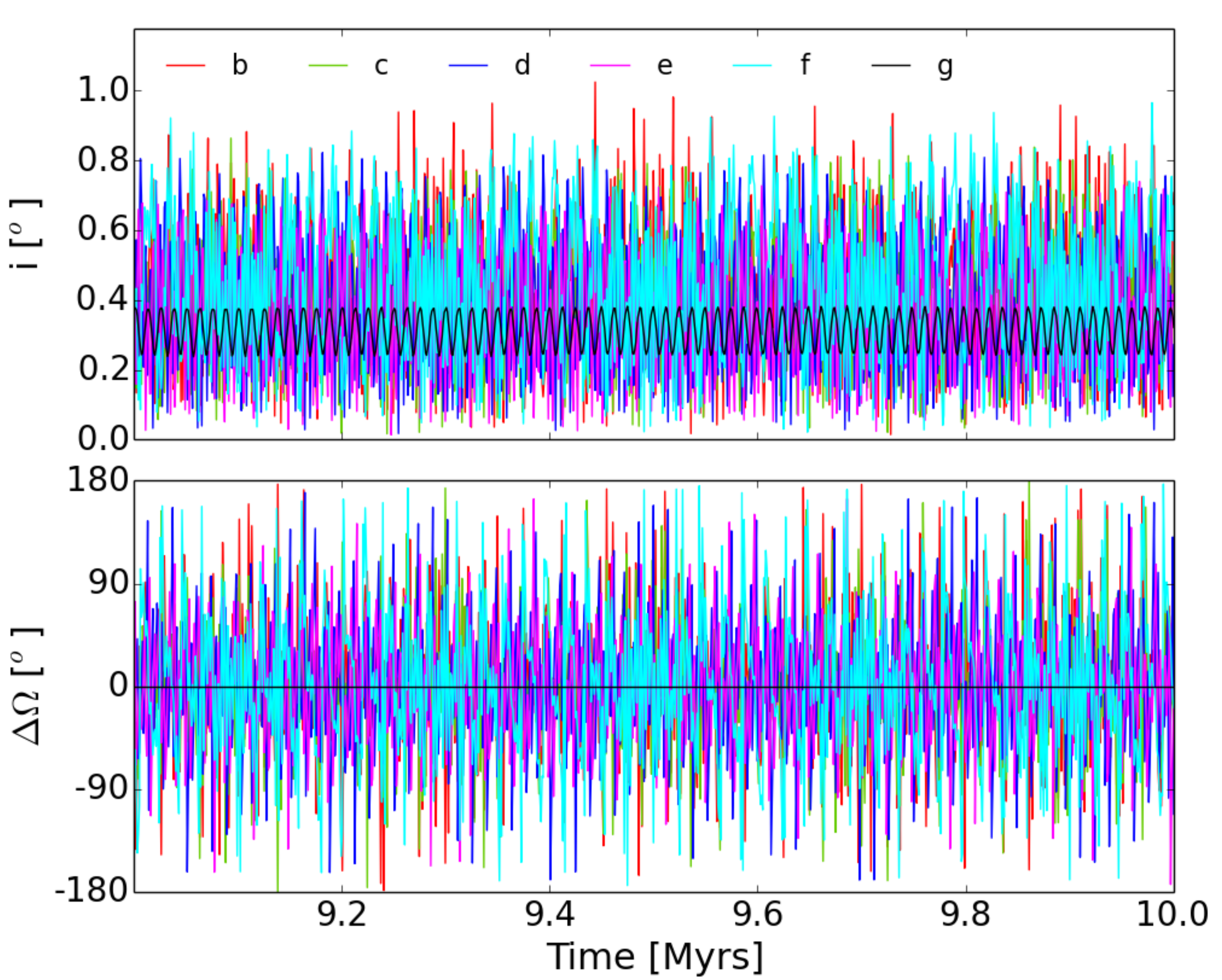}\\
 	\includegraphics[width=0.47\textwidth]{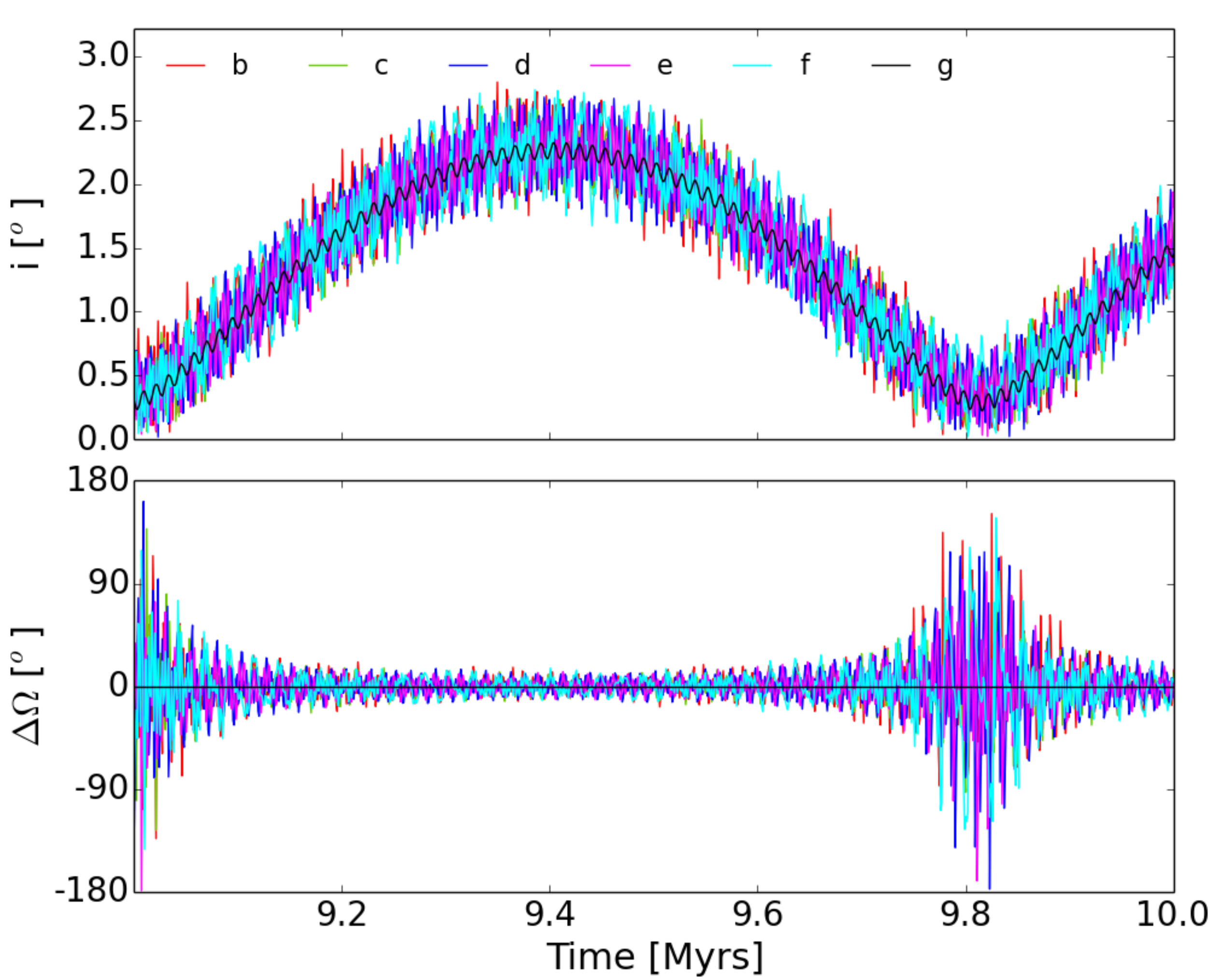}
	\caption{Evolution of the orbital parameters of a stable K11 analogue (top) and K11+ (bottom) system. The top panel shows the 
			inclinations for all of the nominal planets in the system, while the bottom panel shows the longitude of ascending node 
			relative to K11g, the outermost planet.}
	\label{fig:1}
\end{figure}
\begin{figure}[t!]
	\centering
	\includegraphics[width=0.47\textwidth]{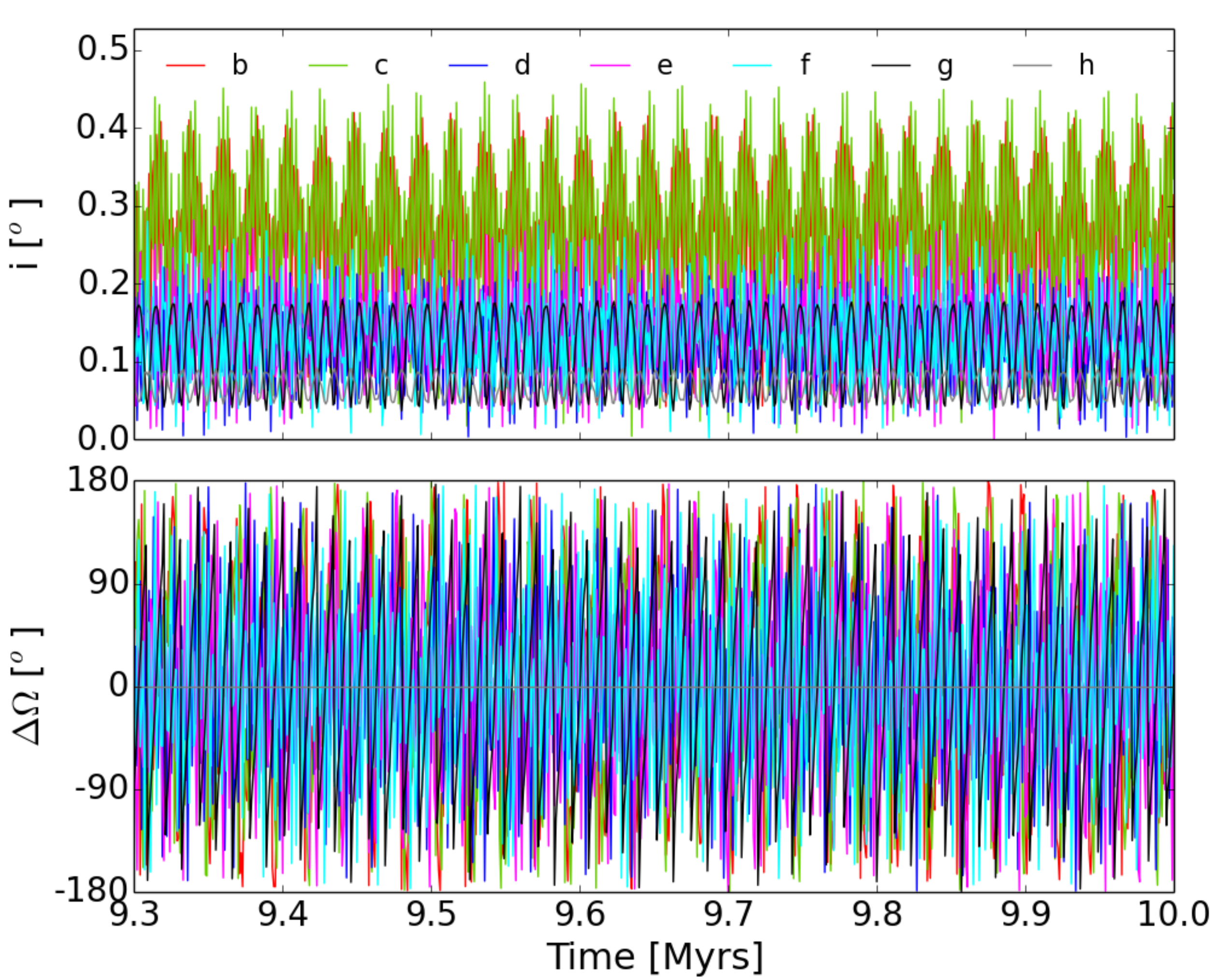} \\
	\includegraphics[width=0.47\textwidth]{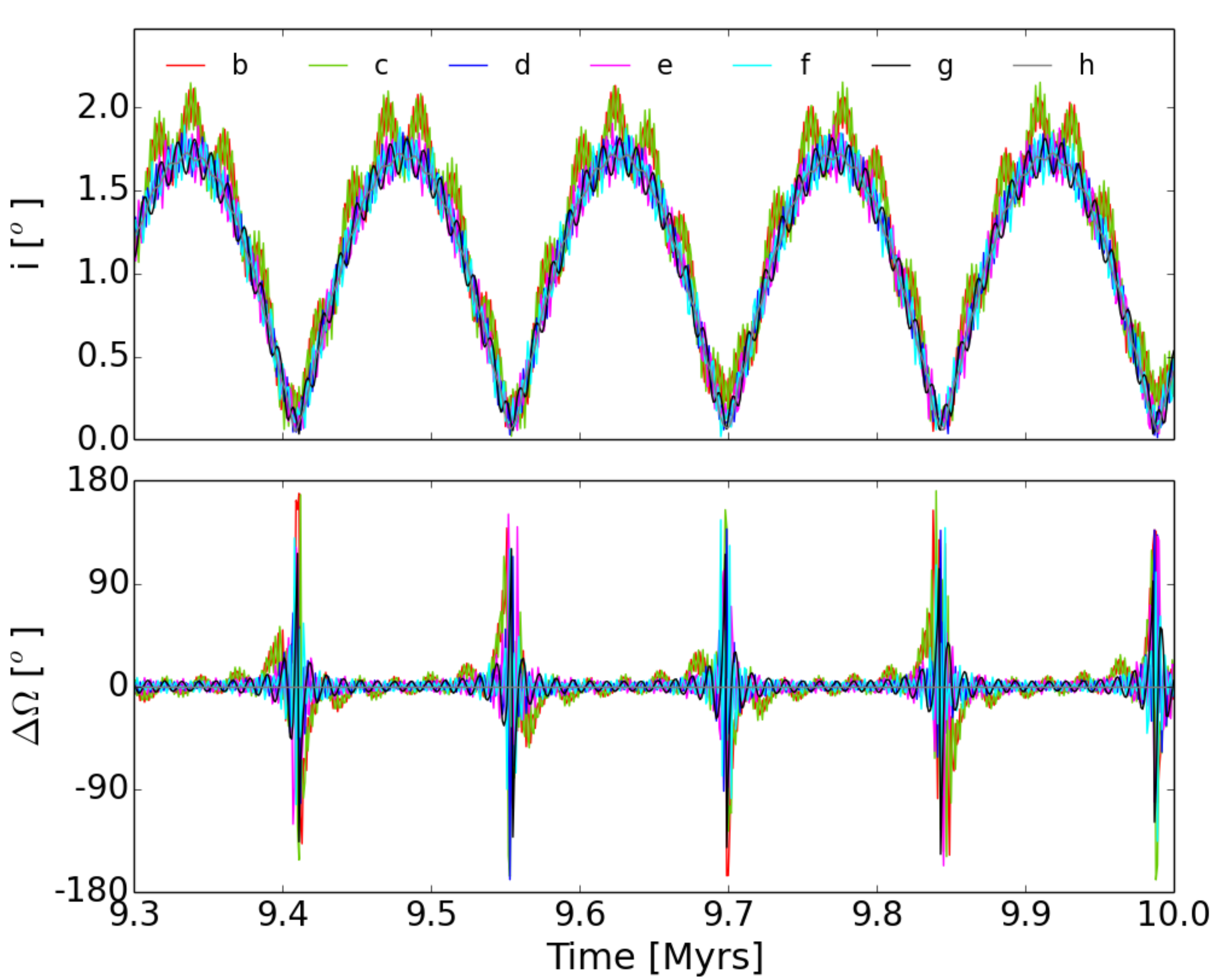}
	\caption{The same orbital parameters in Figure \ref{fig:1} are shown for K90 (top) and K90+ (bottom). Dynamical 
			rigidity is also observed in this system, i.e., all the longitudes of ascending node precess at the same rate. K90b and c 
			(magenta and blue lines) are more tightly coupled to each other than to the rest of the planets.}
	\label{fig:2}
\end{figure}

The simulations of K11 and K90 can now be used as a reference for exploring additional dynamical perturbations. Using the same 
realizations for K11 and K90, we now include a Jupiter analogue in each system as described in section \ref{sec:metho}. We find that 
the stability of the STIPs is not affected by the presence of the single perturber when it is placed at 5.2 au. The same 44\% and 
20\% analogues become unstable in less than 10 Myrs for K11+ and K90+, respectively, which suggest that the instability is driven 
solely by the inner planets, at least for this short time period and for the initial placement of the Jupiter analogue.
\begin{figure*}[hp!]
	\centering
	\includegraphics[width=0.98\textwidth]{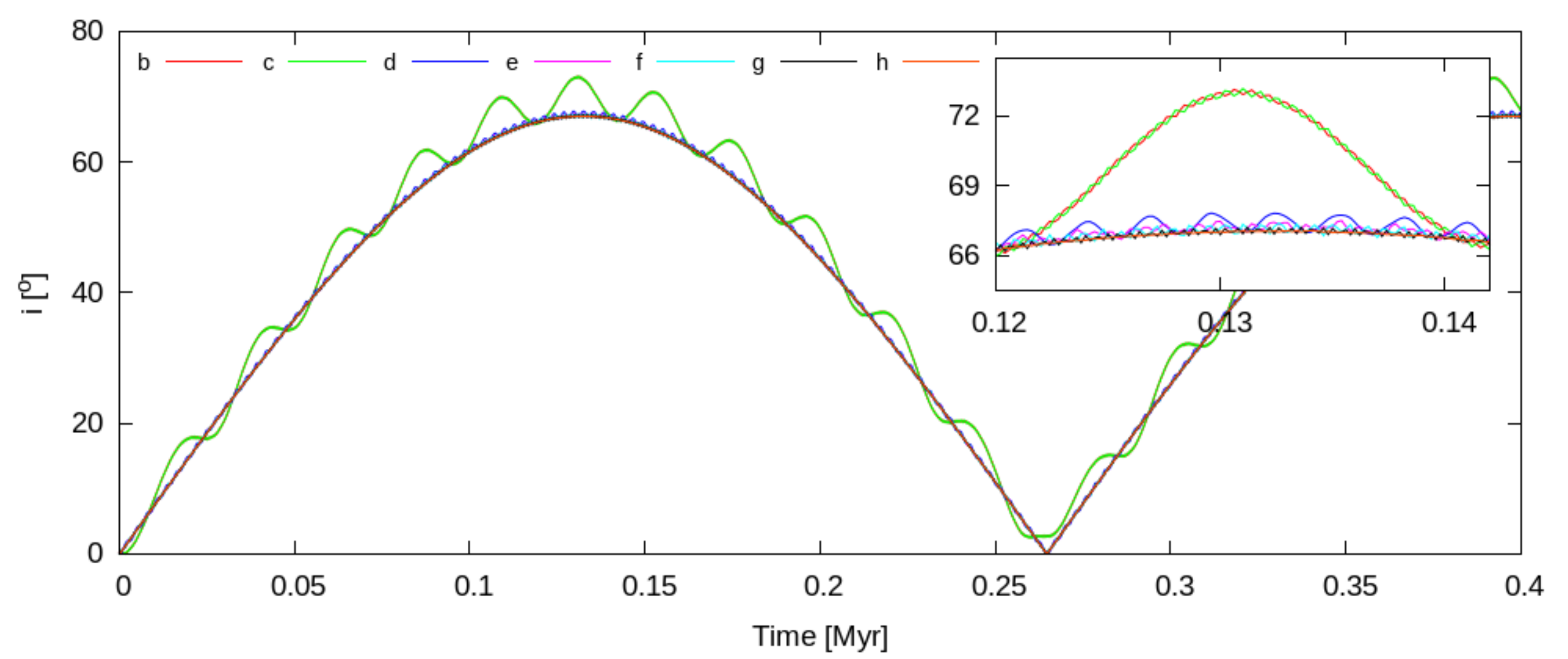}
	\caption{K90+ system with a highly inclined perturber at $50^\circ$ and with $a = 5.2$ au. The dynamical rigidity is evident 
			in the coherent variation of the inclinations. The sub-panel shows a closer look at the evolution of the mutual 
			inclination of the planets. The node of each planet precesses at the same rate, which keeps the mutual inclination
			of the planets small even though the common orbital plane oscillates.}
\label{fig:high_inc}
\end{figure*}
\begin{figure*}[b!]
	\centering
	\includegraphics[width=\textwidth]{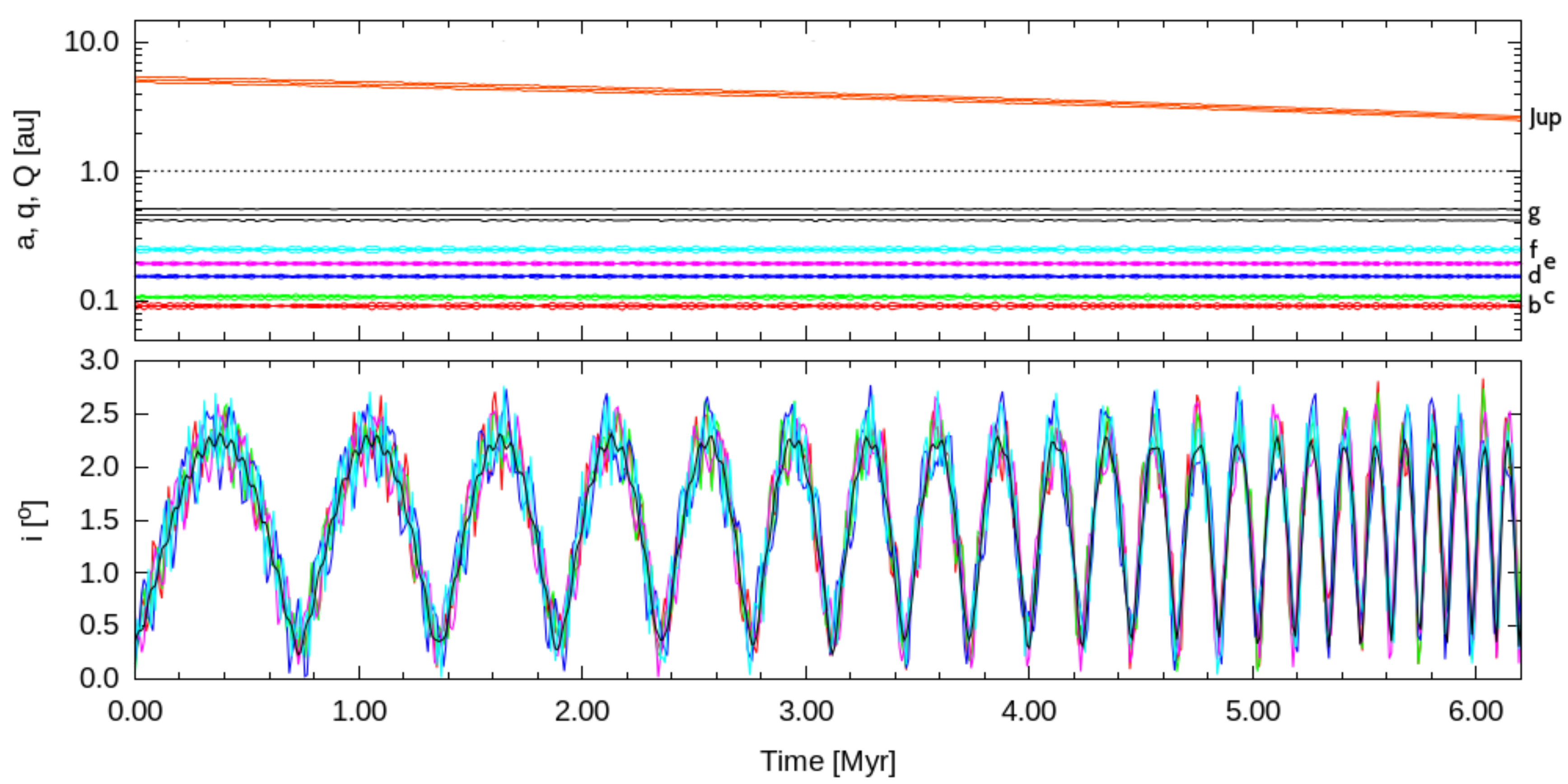}
	\caption{Orbital evolution of K11+ in the presence of a migrating outer perturber. On the top panel, each planet display 3 curves: 
			pericenter, $q$, apocenter, $Q$, and semi-major axis, $a$. The dotted line at 1 au indicates the perturber's location at which a secular interaction is suspected, which in this case, may be driving the instability of the system. In the bottom panel, the orbital inclination, $i$, is shown. 
			The forced migration of the perturber, labeled as Jup, can be observed in orange. Only the first 6.2 Myrs of the 
			simulation is shown. During this time, the system is stable and the precession rate of the common orbital plane increases 
			as a function of the proximity of the perturber. The system becomes unstable when $a_J=0.98$ au (at $t\approx10$ Myrs) }
	\label{fig:K11mig}
\end{figure*}
\begin{figure*}[t!]
	\centering
	\includegraphics[width=\textwidth]{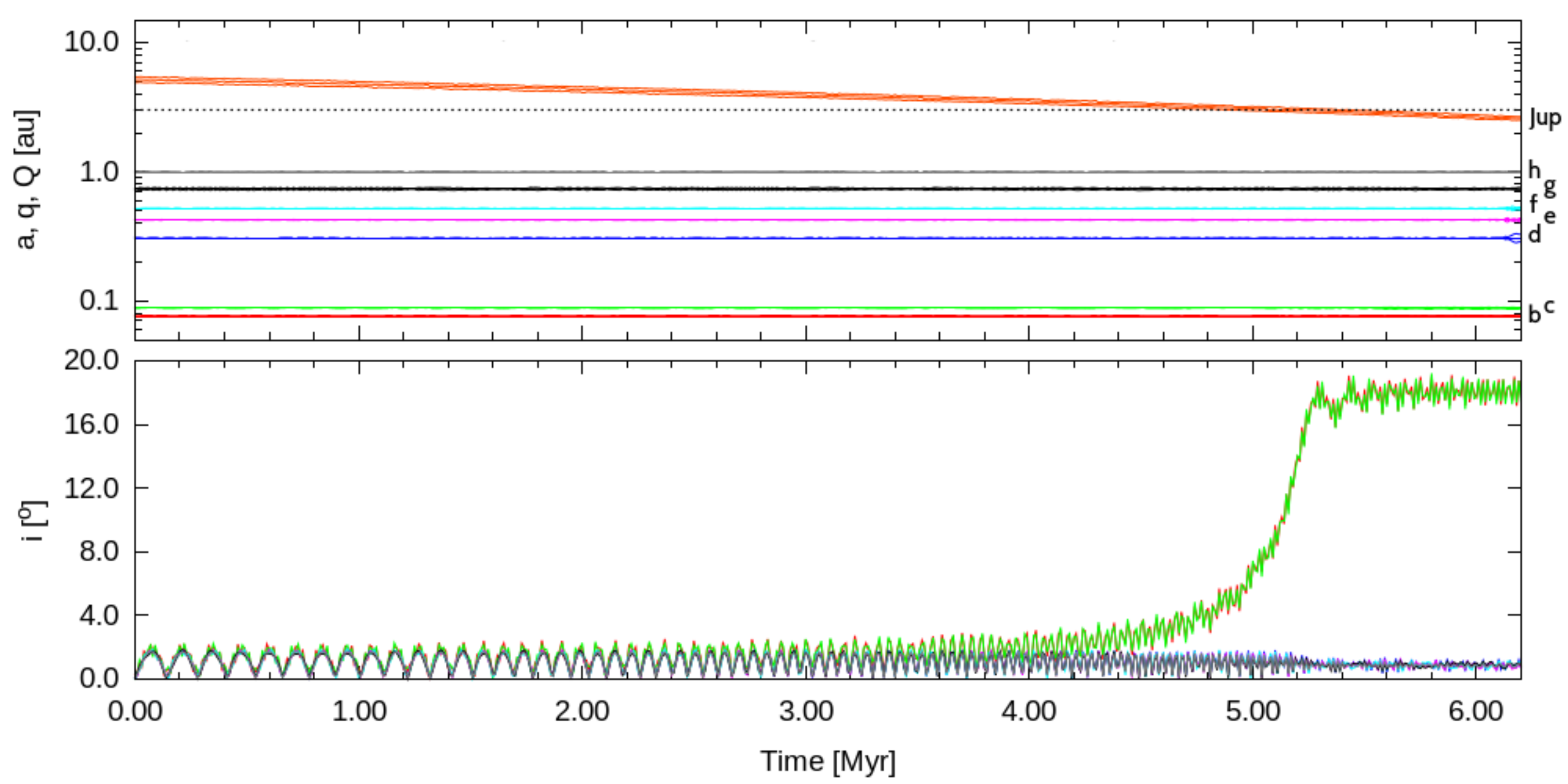}
	\caption{Orbital evolution of K90+ in the presence of a migrating outer perturber. The orbital elements displayed are the same 
			that those in Figure \ref{fig:K11mig}. The precession behavior of K90+ is similar to K11+, with K90+ becoming unstable 
			when $a_J = 2.5$ au (at $t \approx 6.3$ Myrs). However, between perturber orbital distances of $a_J \sim 2.5$ au and 3.2 au, K90b and c become excited together onto a second plane of higher inclination relative to their original orbits. The dotted line in the top panel indicates the perturber's position at which the inclinations of K90b and K90c become about 10$^\circ$ larger than the inclinations of the other planets. }
	\label{fig:K90mig}
\end{figure*}

While the stability of the STIPs is unaffected, the systems do respond to the presence of the outer perturber. This is highlighted 
in Figures \ref{fig:1} and \ref{fig:2}, which show the inclination evolution for the planets in stable realizations of K11 and K90, 
respectively, with and without the presence of the Jupiter analogue. In each case, the planets exhibit variation among their 
inclinations with time. When the perturber is included, however, there is an additional large-scale variation in all of the orbital 
inclinations, i.e., each planet's inclination oscillates with respect to a common orbital plane while the orientation of the 
shared orbital plane changes. The presence of the perturber precesses the longitudes of ascending node of all the inner planets at the 
same rate, which is also shown in Figures \ref{fig:1} and \ref{fig:2} (with $\Delta \Omega$). Such ``dynamical rigidity'' has been 
seen in simulations of other systems when perturbers are introduced \citep{Kaib11} at high inclination, and has been further explored 
analytically \citep{boue14a,boue14b}. Several test simulations were run with the Jupiter analogue at high inclination ($50^\circ$), 
which showed that the longitudes of ascending node remained locked, with the STIP undergoing large and coherent inclination variations 
(Fig. \ref{fig:high_inc}).

We have thus far only considered one location for a perturber, arbitrarily introduced at 5.2 au. We now explore a wider range of 
semi-major axes by forcing the perturber to migrate inwards from 5.2 au to $\sim$ 1 au. First, the overall behavior of the STIPs is 
unchanged by simply moving the Jupiter analogue inwards. Figures \ref{fig:K11mig} and \ref{fig:K90mig}, for example, show that the 
systems continue to exhibit coherent changes in the orbital plane, but with an increasing precession rate of that plane as the Jupiter 
analogue approaches the inner system. However, for certain system configurations, the outcomes can be very different. In most cases, 
this appears to be due to shifts in the secular frequencies. For example, eccentricity resonances can drive the STIP towards 
instability, causing planetary collisions (as occurs for K11). An inclination resonance can force a planet (or planets) out of the 
original common orbital plane, effectively creating multiple orbital planes in a stable system (as occurs in K90). Specifically, runs 
with an initially stable K11+ develop an instability if the perturber reaches a semi-major axis of $\sim 1.0$ au, which places 
the perturber in a 3:1 near-MMR with K11g. This commensurability could be causing the instability in K11+, but the secular resonances 
could also have a strong contribution to the destabilization of K11+, as will be discussed shortly. At this time, 59\% of the 
unstable K11+ realizations result in either K11b crossing the orbit of K11c followed by K11f crossing K11e's orbit or vice versa\footnote{K11f crossing the orbit of K11e first and then K11b crossing K11c's orbit.}. In K90+, when $a_J \lesssim 3.2$ au, K90b and K90c evolve together away from the rest of the planets by about $16^\circ$ in inclination, with the system remaining stable.

\begin{figure*}[p]
    \centering
	\includegraphics[width=\textwidth]{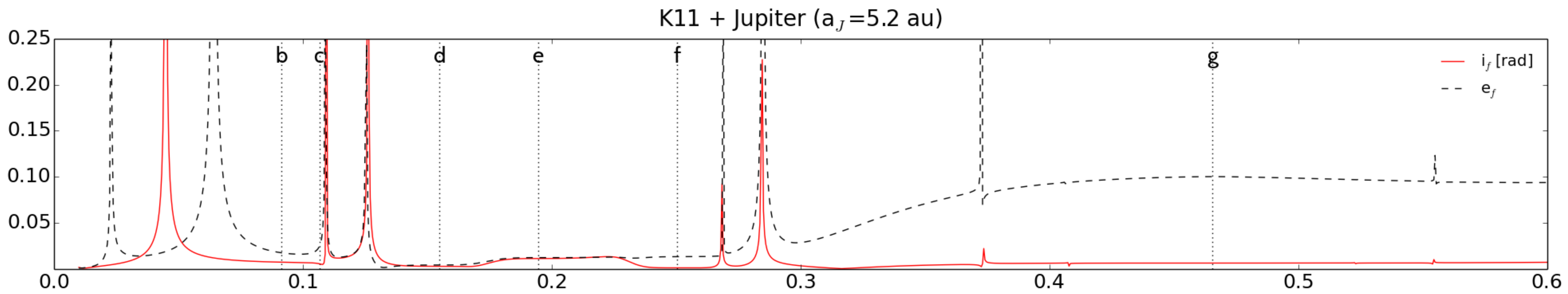}
	\includegraphics[width=\textwidth]{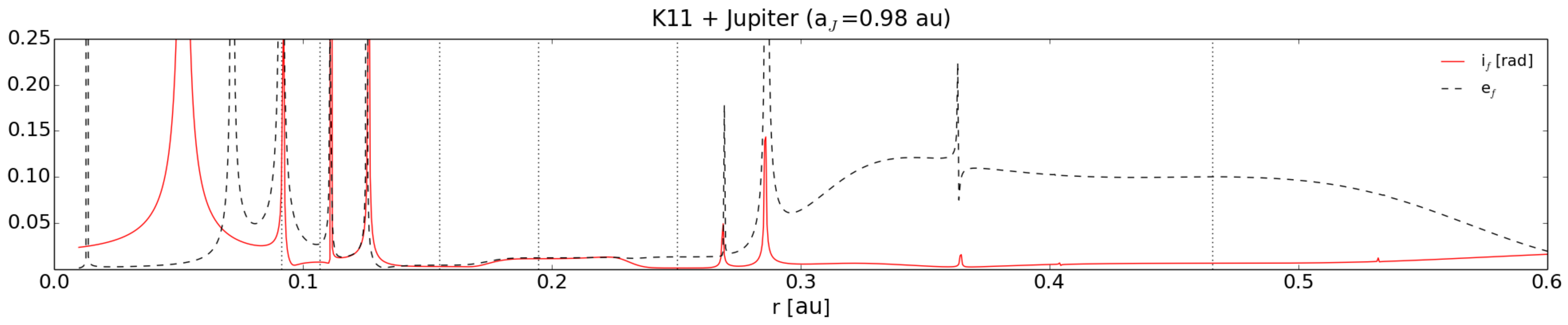}
	\caption{Secular map of the forced inclination (red solid line) and eccentricity (black dashed line) of K11 in the presence of a
			Jupiter-like perturber and excluding K11b and K11c in the calculations. Two different semi-major axes of the perturber are
			shown in the top and bottom panels. The locations of the inner planets are shown by vertical dotted lines. \textit{Top 
			panel:} the gas giant is at $a_J = 5.2$ au, and none of the resonances coincides with the location of the inner planets. 
			In contrast, when $a_J = 0.98$ au (bottom panel), inclination and eccentricity resonances are located at the position of 
			K11b. In this case the eccentric resonance appears to contribute to destabilizing the system.}
	\label{fig:secK11}
\end{figure*}
\begin{figure*}[h!]
    \centering
	\includegraphics[width=\textwidth]{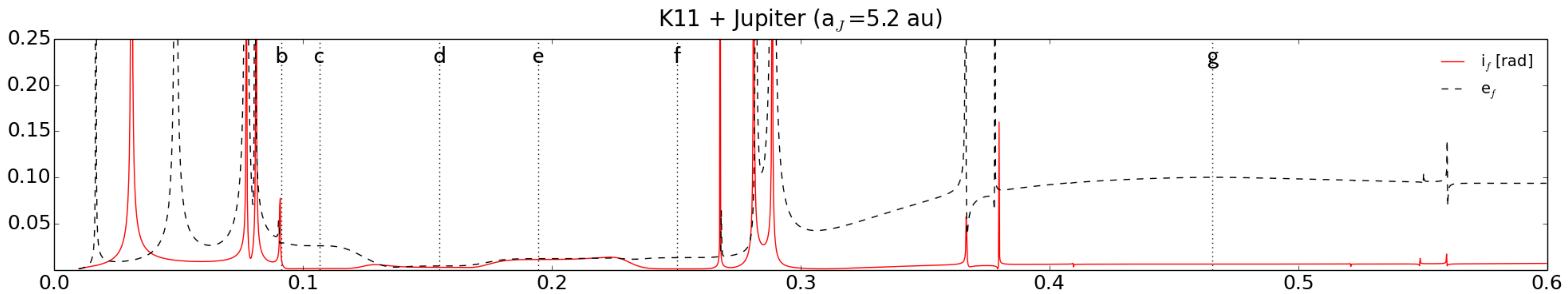}
	\includegraphics[width=\textwidth]{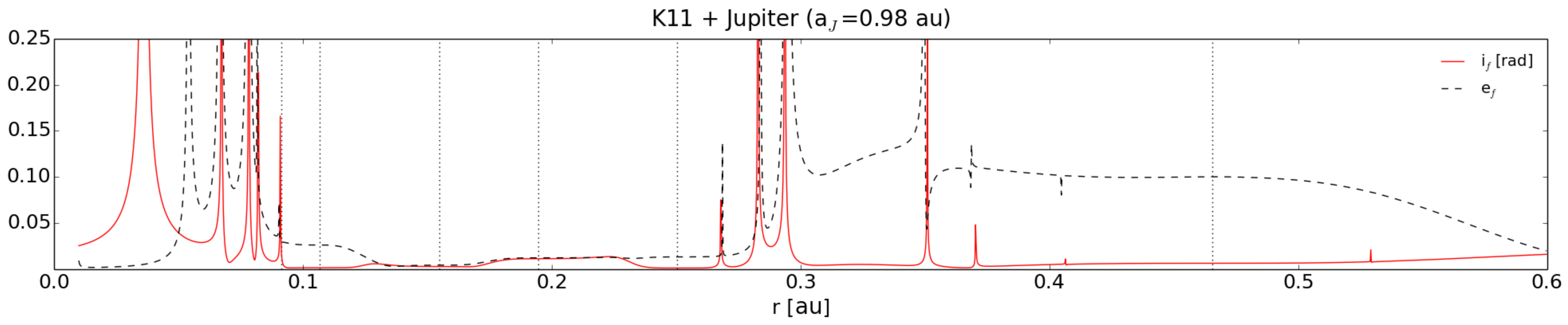}
	\caption{Similar to Fig.~\ref{fig:secK11}, but excluding only K11b. As before, the panels show the results for two different 
			perturber locations. There are multiple inclination and eccentricity resonances just interior to K11b's position. Due to 
			the tight coupling between K11b and K11c, Fig.~\ref{fig:secK11} might better reflect the onset of instability. 
			Nevertheless, these panels highlight the richness of the secular structure of the inner system. }
	\label{fig:secK11-2}
\end{figure*}
\begin{figure*}[t!]
    \centering
	\includegraphics[width=\textwidth]{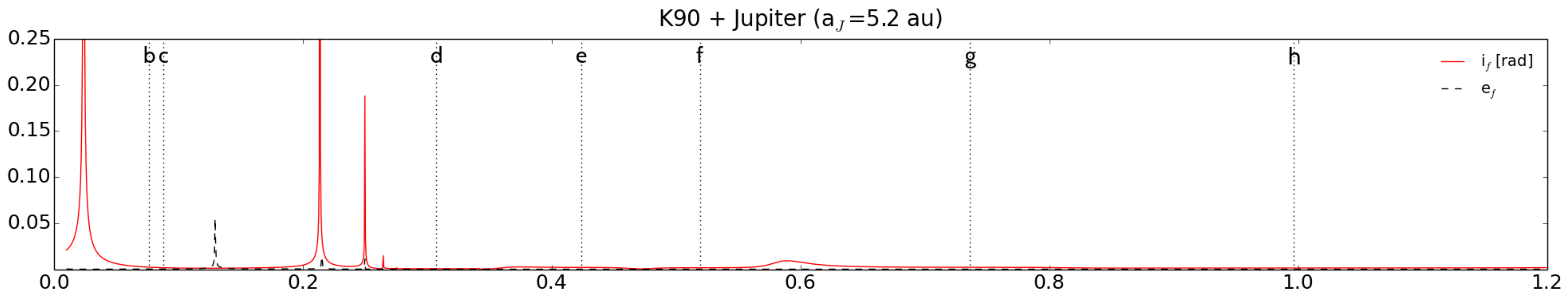}
	\includegraphics[width=\textwidth]{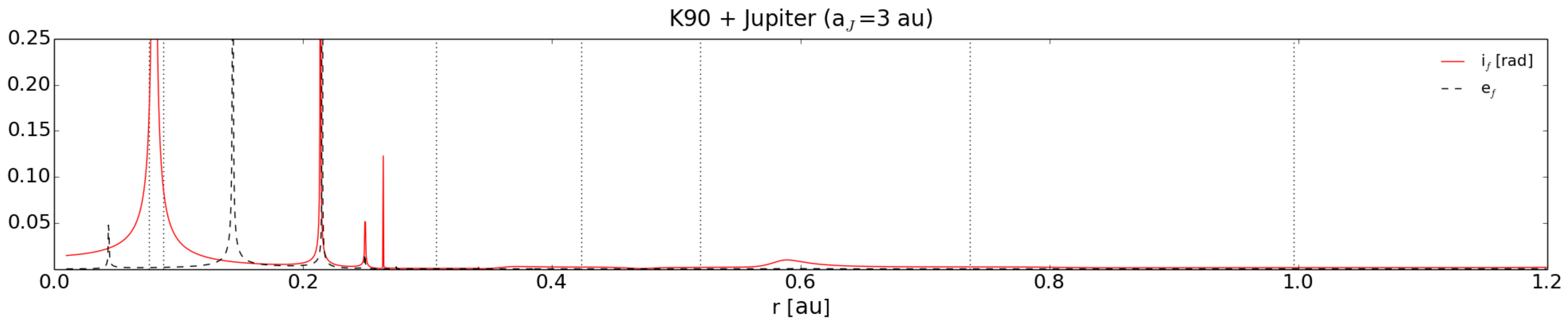}
	\caption{Secular map of the forced inclination and eccentricity of K90 in the presence of a Jupiter-like perturber. \textit{Top 
			panel:} the perturber is at $a_J = 5.2$ au, and none of the resonances coincides with the location of the inner planets. 
			When we consider $a_J = 3.0$ au (bottom panel), a wide inclination resonance is located at the position of K90b and K90c. 
			This resonance increases the inclination of both K90b and K90c without affecting the stability of the system.}
	\label{fig:secK90}
\end{figure*}

The behavior of both K11+ (Fig. \ref{fig:secK11} and \ref{fig:secK11-2}) and K90+ (Fig. \ref{fig:secK90}) can be understood by looking 
at their forced eccentricities and inclinations. As the Jupiter analogue moves inwards, the locations of the innermost inclination and 
eccentricity resonances move outwards in the STIP, eventually crossing the innermost planets. For K11+, an inclination and 
eccentricity resonance overlap the semi-major axis of K11b when $a_J = 0.98$ au, which is when instability occurs in the 
corresponding simulations (Fig. \ref{fig:secK11}). For this analysis, K11b and K11c were removed from the secular theory calculation. 

In the case of K90+, when the perturber is at $a_J \approx 3.3$ au the two innermost planets appear to be trapped in an inclination 
secular resonance as determined by secular theory (with planets K90b and K90c removed, Fig. \ref{fig:secK90}). Together, Figures 
\ref{fig:K90mig} and \ref{fig:secK90} suggest that the separation of the K90+ system into two distinct orbital planes (for $a_J 
\lesssim 3.2$ au) is due to an inclination resonance overlapping the K90b and K90c positions. This secular resonance excites the 
inclination of both planets, which are strongly coupled, and its strength increases as the perturbing planet's distance to 
K90 decreases. This second orbital plane for K90b and K90c can acquire a maximum inclination $i_{bc} \sim 18^o$ if $a_J \sim 3.0$ au. 
While there is also an inclination overlapping the location of K11b in K11+ (when $a_j\approx 0.98$ au), there is also an 
eccentricity resonance present (Fig. \ref{fig:secK11}) possibly leading to instability before any large inclination changes could 
occur.

As already discussed, removing planets b and c from both K11+ and K90+ allow us to explore whether there are forced inclination and 
eccentricities at their locations. This assumes that the planets are massless, which is not true. As such, the method is not 
guaranteed to highlight resonances, although the correspondence with the N-body simulations suggest the approximation is valid in this 
case. The calculation nonetheless reduces the complexity of the secular structure by removing frequencies. To highlight this, we show 
in Figure \ref{fig:secK11-2} a secular map in which we only remove K11b, allowing K11c to contribute to the secular model. An 
inclination and eccentricity resonance is present at $r \approx 0.09$ au, just inside K11b. The position of this resonance is fixed 
regardless of the perturber's proximity to the inner system, suggesting that the location of this resonance is due to the K11+ inner 
planets. There are additional resonances that are at even smaller semi-major axes, which are affected by the perturber's location. As 
the perturber moves inwards, this inner resonance structure moves outwards and the resonance wings overlap. At face value, Figure 
\ref{fig:secK11-2} suggests that K11b might not overlap a secular resonance. We will later show that K11b and K11c are strongly 
coupled and Figure \ref{fig:secK11} might better reflect the outer system's influence on K11b and K11c. Nevertheless, Figure 
\ref{fig:secK11-2} shows the proximity of these resonances to K11b, which could explain the observed instability of the system and its 
apparent \textit{fine tuning even when the perturber is absent}. A small perturbation in the configuration of planets could cause K11b 
to enter a secular eccentricity resonance and collide with K11c, the most common outcome in our simulations that become unstable.
\begin{figure*}[t!]
	\centering
	\includegraphics[width=0.495\textwidth]{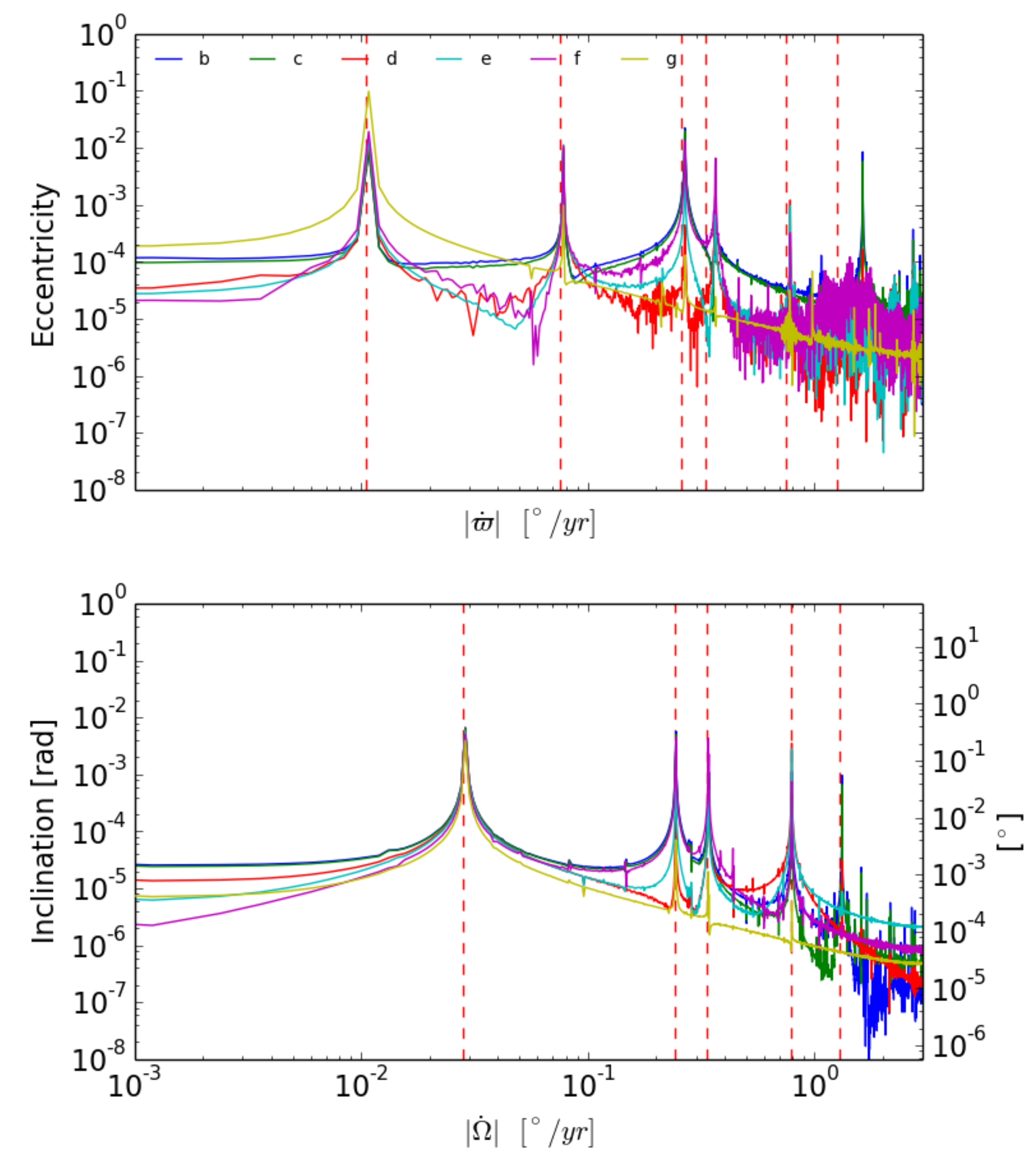}
	\includegraphics[width=0.495\textwidth]{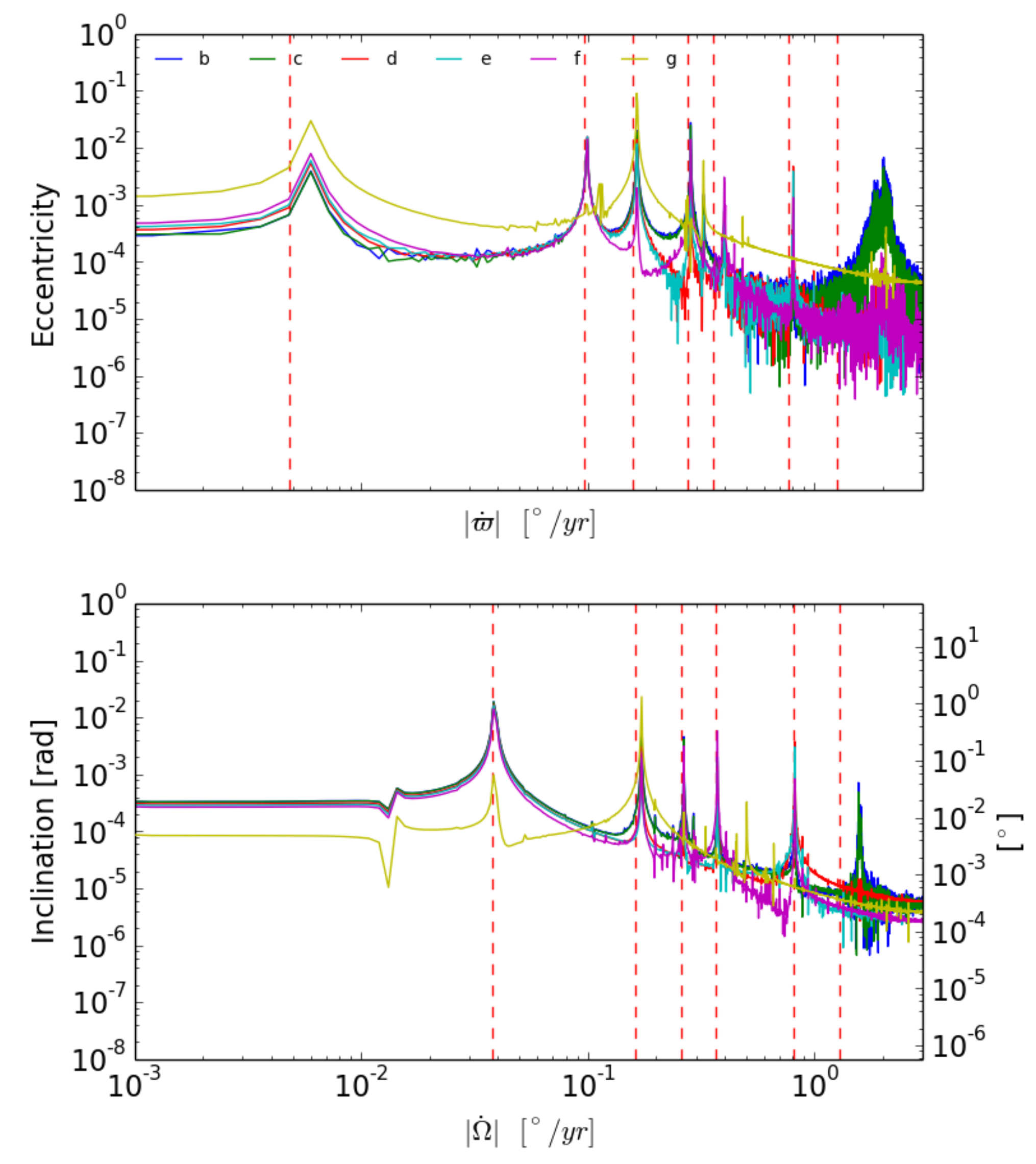}
	\caption{Comparison of the precession frequencies of K11 (left) and K11+ (right) with their secular eigenfrequencies (red dashed 
			lines). Each of the colored curves represent a planet, as indicated in the legend. \textit{Top panels:} K11 and K11+ 
			apsidal precession spectra, where at least two of the principal apsidal precession frequencies are offset from the nearest 
			eigenfrequency. \textit{Bottom panels:} Nodal precession spectra, there is a closer alignment of the principal precession 
			frequencies with the eigenfrequencies than in the apsidal case, particularly for K11. The frequency shift relative to secular theory indicates that an unaccounted secular non-linear resonance, in eccentricity, is present in K11+.}
	\label{fig:precK11}
\end{figure*}

\begin{figure*}[t!]
	\centering
	\includegraphics[width=0.495\textwidth]{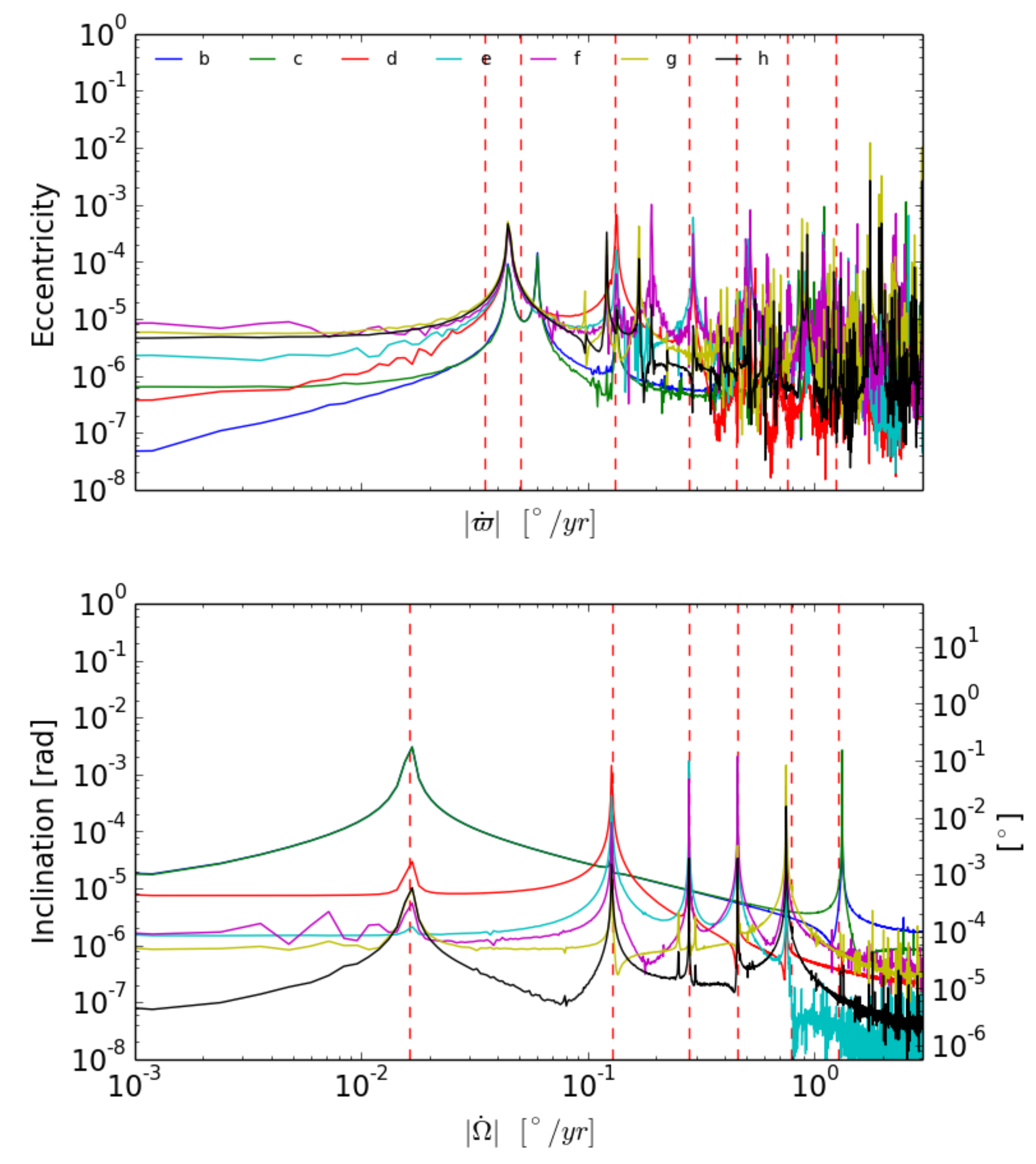}
	\includegraphics[width=0.495\textwidth]{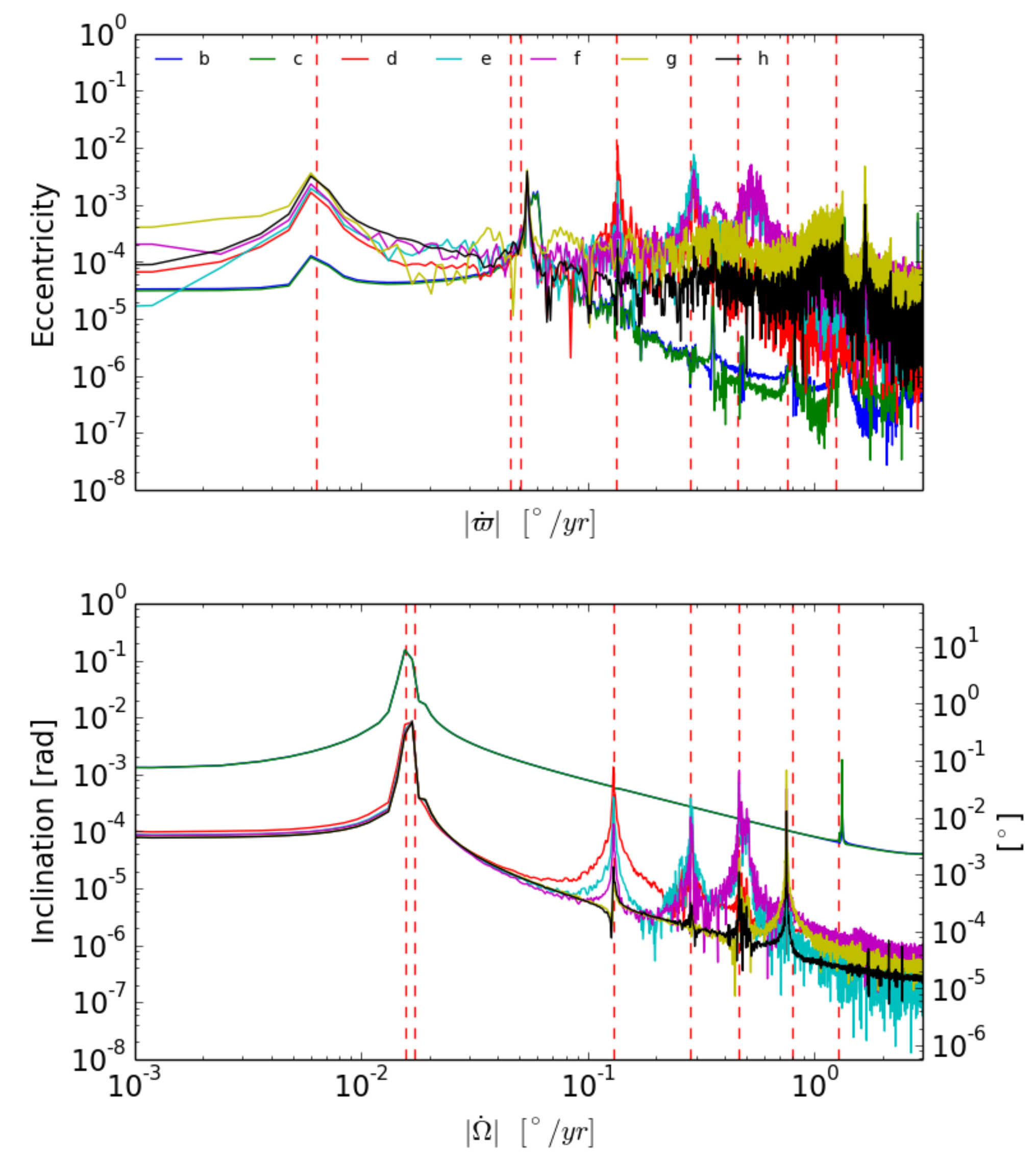}
	\caption{Comparison of the precession frequencies of K90 (left) and K90+ (right) with their secular eigenfrequencies (red dashed 
			lines). The notation used in Figure \ref{fig:precK11} is adopted here. In the apsidal spectra, of both K90 and K90+, 
			there is aliasing at higher $\dot{\varpi}$ (short periods) and the characteristic precession frequencies miss most the 
			estimated eigenfrequencies. In contrast, the characteristic nodal precession frequencies are in close alignment with the 
			eigenfrequencies and the aliasing is minor to moderate.}
	\label{fig:precK90}
\end{figure*}

We investigate the secular structure of the systems further by calculating synthetic secular frequencies (precession spectra) with and 
without the perturber. The precession spectra are determined directly from the N-body calculations, using 0.3 Myrs of output with a 
sample time of 60 years. In K11+ and K90+, the perturber was set to $a_J = 0.98$ au and $a_J = 3.0$ au, respectively. The comparison 
of the precession spectra with apsidal and nodal secular eigenfrequencies is shown in Figures \ref{fig:precK11} and \ref{fig:precK90}. 
As expected, the value and number of eigenfrequencies change from K11 to K11+ and K90 to K90+, respectively, due to the additional 
planet in the perturbed systems, with a greater displacement for the apsidal eigenfrequencies.
The amplitude of the global Fourier spectra, in both eccentricity and inclination, for K11+ and K90+ is higher than in the unperturbed 
analogues; there is about one order of magnitude difference in K11+ and two to three in K90+. The nodal eigenfrequencies show very 
good agreement with the synthetic spectra for both K11(+) and K90(+), although there are some deviations. The apsidal precession 
frequencies of both systems exhibit stronger disagreement between the synthetic spectra and the secular theory, but the features in 
the Fourier spectra remain recognizable, particularly for K11(+). The misalignment of the principal apsidal frequencies relative to 
the estimated secular eigenfrequencies might indicate that there are strong interactions among the planets, which are neglected in the 
secular calculation. 
This could include non-linear secular resonances and near mean motion resonances (MMR). \citet{Malhotra89} showed that the $e-\varpi$ 
eigenfrequencies are significantly shifted if a near first order MMR is present in the system because the averaged effect of the term 
$e \, \cos(j\lambda - (j+1)\lambda'+ \varpi)$ in the secular expansion is non-zero. There is a weaker effect of such a near-resonant 
term on $i-\Omega$ because the nodal term is associated with $i^2$. The MMR can also add eigenfrequencies to the system, as occurs in 
the Solar System due to the near-MMR between Jupiter and Saturn \citep{Brouwer50}.
\begin{figure}[t!]
	\centering
	\includegraphics[width=.5\textwidth]{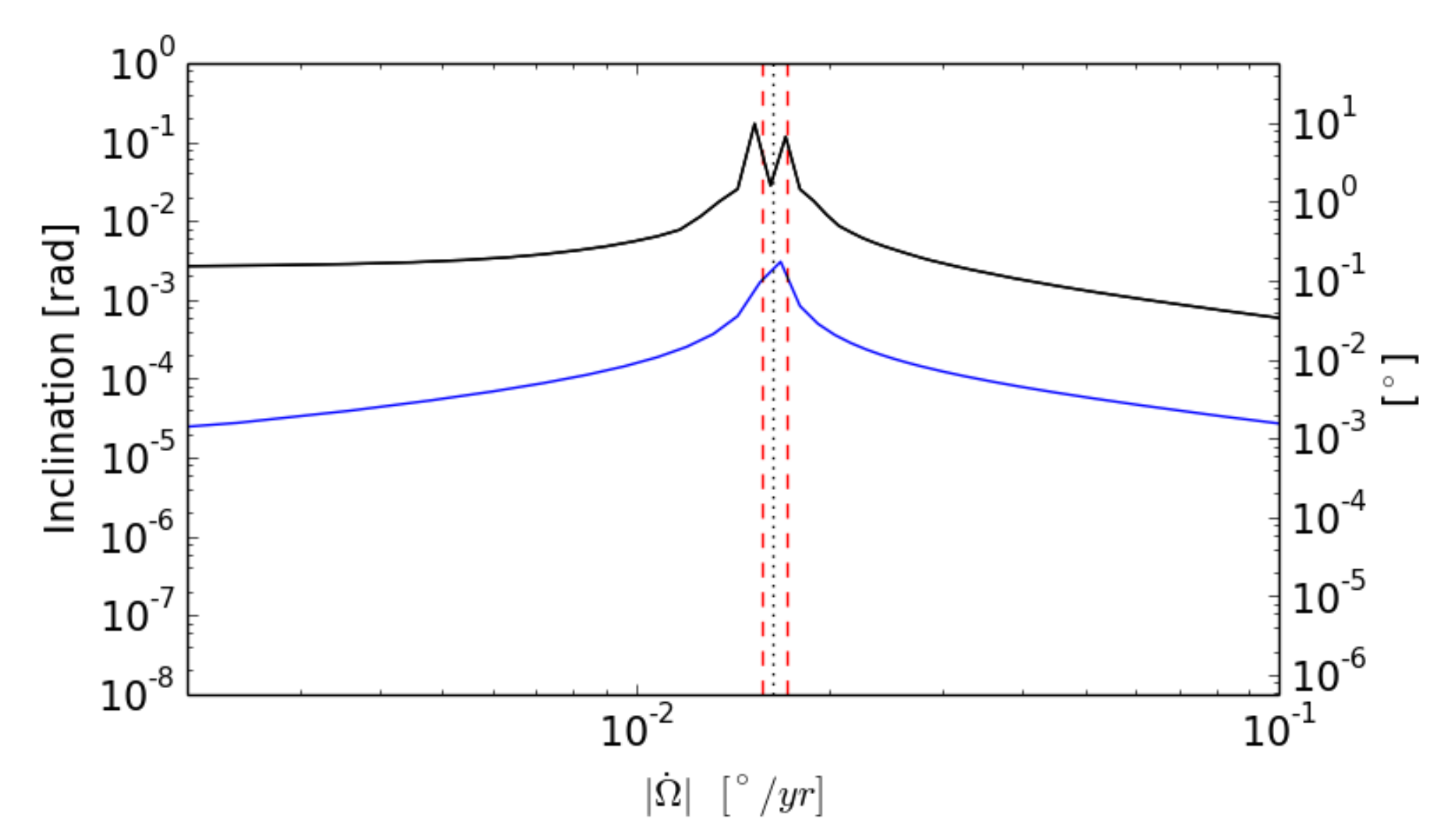}
	\caption{Zoom-in of the nodal principal component of planet b in K90 (blue line) and K90+ (black solid line). In this case, K90+ 
			was evolved for 0.4 Myrs, which increased the resolution of the Fourier transform. The K90+ dominant peak in Fig. 
			\ref{fig:precK90} splits into two peaks when the resolution is increased. The vertical red dashed-lines represent the 
			eigenfrequencies for K90+ in the given range, while the black dotted vertical line corresponds to the eigenfrequency for 
			K90 in that range.}
	\label{fig:zK90b(+)}
\end{figure}

The main precession period of each planet, in K11(+) and K90(+), and the corresponding amplitude are displayed in Table 
\ref{tab:mainP}. In the four systems, the two innermost planets have the same main precession period ($\dot{\varpi}$ and 
$\dot{\Omega}$) and their precession spectra are almost indistinguishable, confirming the strong coupling of these planetary pairs. 
Furthermore, in K11, the nodes of the six planets precesses with the same period, suggesting that K11g is dominating the nodal 
precession, which causes a dynamical rigidity in the orbital plane. On the other hand, there are two distinct pericenter precession 
periods, in which K11b and K11c have an apsidal period of $P_{K11b-c} \simeq 1346$ yr and K11d through K11g have periods of 
$P_{K11d-g} \simeq 33340$ yr. The perturber in K11+ changes the frequencies and decouples K11g; K11g's node precesses at the same rate 
as the perturber, which is $\sim 4$ times faster than the rest of planets in K11+.
In contrast to K11, the main nodal precession periods of K90 do not suggest an intrinsic dynamical rigidity, although there is strong 
coupling between K90b and K90c, as well as K90g and K90h. However in K90+, there are two distinct nodal precession periods that 
are different by only 7\%. These nearly overlapping eigenfrequencies (Fig. \ref{fig:precK90}) suggest a secular resonance is causing 
the large change in inclinations for K90b and K90c (Fig. \ref{fig:K90mig}). The amplitude of the broad frequency peak seen in 
Fig.~\ref{fig:precK90} is dominant by one order of magnitude or more. When the K90+ simulation is allowed to evolve to 0.4 Myrs, it is 
possible to identify the two close frequency components in the dominant broad peak of Figure \ref{fig:precK90}, as expected from 
secular theory for K90+. These are shown in Figure \ref{fig:zK90b(+)}, with the peaks corresponding to $P_\mathrm{node} \simeq 23100$ 
yr and $P_\mathrm{node} \simeq 21400$ yr, which straddle the eigenfrequency associated with K90 in that range. The amplitude of these 
two frequencies is interchanged when the spectra is calculated from a subsample. The whole precession spectra of the longer N-body 
integration is not shown in Fig. \ref{fig:zK90b(+)} due to severe aliasing at large frequencies.

\begin{deluxetable*}{ccccccccc}[th!]
\tablecolumns{9}
\tablecaption{Main precession period and corresponding amplitude for each planet in K11, K11+, K90, and K90+. \label{tab:mainP}}
\tablehead{ \colhead{Planet} & \mc{2}{c}{K11}     & \mc{2}{c}{K11+}   & \mc{2}{c}{K90}   & \mc{2}{c}{K90+}   \\ \hline
\colhead{}  & \colhead{$P_\mathrm{ap}$}  & \colhead{$e$} & \colhead{$P_\mathrm{ap}$} & \colhead{$e$} & \colhead{$P_\mathrm{ap}$}   & \colhead{$e$}        & \colhead{$P_\mathrm{ap}$}   & \colhead{$e$} \\
\colhead{}  & \colhead{(yrs)}  & \colhead{} & \colhead{(yrs)} & \colhead{} & \colhead{(yrs)}   & \colhead{} & \colhead{(yrs)}  & \colhead{} 
}
\startdata
		b & 1346  & 0.0225 & 1266  & 0.0276 & 104  & 0.0008 & 6667  & 0.0028 \\
		c & 1346  & 0.0201 & 1266  & 0.0246 & 104  & 0.0011 & 6667  & 0.0026 \\
		d & 33340 & 0.0131 & 3615  & 0.0158 & 2703 & 0.0007 & 2655  & 0.0110 \\
		e & 33340 & 0.0146 & 3615  & 0.0157 & 138  & 0.0007 & 1230  & 0.0078 \\
		f & 33340 & 0.0194 & 3615  & 0.0142 & 205  & 0.0012 & 695   & 0.0050 \\
		g & 33340 & 0.0988 & 2190  & 0.0908 & 205  & 0.0121 & 216   & 0.0047 \\
		h & ...   & ...    & ...   & ...    & 205  & 0.0026 & 6667  & 0.0039 \\
		J & ...   & ...    & 60012 & 0.0436 & ...  & ...    & 60006 & 0.0398 \\ 
		\hline
		  & $P_\mathrm{node}$	& $i$  & $P_\mathrm{node}$	& $i$  & $P_\mathrm{node}$	   & $i$  & $P_\mathrm{node}$	 & $i$  \\ 
		  & (yrs)  & ($^\circ$) & (yrs) & ($^\circ$) & (yrs) & ($^\circ$) & (yrs) & ($^\circ$) \\
		\hline
		b & 12503 & 0.3815 & 9377 & 1.0840 & 21431 & 0.1738 & 23079 & 8.8099 \\
		c & 12503 & 0.3777 & 9377 & 1.0728 & 21431 & 0.1732 & 23079 & 8.7856 \\
		d & 12503 & 0.3492 & 9377 & 0.9740 & 2831  & 0.0831 & 21431 & 0.4781 \\
		e & 12503 & 0.3374 & 9377 & 0.9362 & 1288  & 0.0988 & 21431 & 0.4999 \\
		f & 12503 & 0.2893 & 9377 & 0.8085 & 785   & 0.1174 & 21431 & 0.4982 \\
		g & 12503 & 0.2253 & 2084 & 1.3211 & 480   & 0.0864 & 21431 & 0.4880 \\
		h & ...   & ...    & ...  & ...    & 480   & 0.0164 & 21431 & 0.4829 \\
		J & ...   & ...    & 2084 & 0.0525 & ...   & ...    & 21431 & 0.3145 \\
\enddata
\tablecomments{$P_\mathrm{ap}$ and $P_\mathrm{node}$ denote the apsidal and nodal period, respectively. The values of $e$ and $i$ 
			provided here are the maximum amplitudes from Figures \ref{fig:precK11} and \ref{fig:precK90} for each planet, and 
			whenever pertinent, the perturber. The apparent switch of the nodal precession from planets K90b and K90c to K90+d through 
			K90+h is an artifact of the Fourier frequency spacing, along with a modest change in the eigenfrequency structure. 
			Comparing the tabulated periods among the planets in a system helps to highlight which planets exhibit tight coupling.}
\end{deluxetable*}

The identification of the frequency components in the apsidal precession spectra of K90 and K90+ (top panels in Fig. 
\ref{fig:precK90}) is also complicated due to aliasing. Even so, we can discern a substantial misalignment of the main frequency 
components compared with the nearest eigenfrequency, as mentioned above. In K90, there is a double-peaked feature corresponding to 
K90b and K90c at $\dot{\varpi}\sim 0.05^\circ/$yr. This is absent in the spectra of the other planets, which could originate from a 
non-linear secular resonance or from an MMR.\\
\section{Discussion}
\label{sec:discussion}

The discrepancies between the frequency components of the precession spectra of K11 and K90 and their secular eigenfrequencies 
demonstrate that either additional physics should be included in the secular theory or that we need to consider a higher order 
expansion in $e$, $i$, and mass. Some of the physical effects that we know we are neglecting in the secular code include GR and MMRs. 
The GR contribution to the precession rates is small; it only affects the apsidal precession frequencies of the synthetic spectra by 
$\lesssim 2 \%$, with K90 being the most affected. GR does not shift the nodal spectra in either system. The MMR, depending on the 
order, can affect linear terms in eccentricity and inclination (first order). 
\begin{figure*}[ht!]
	\centering
	\includegraphics[width=\textwidth]{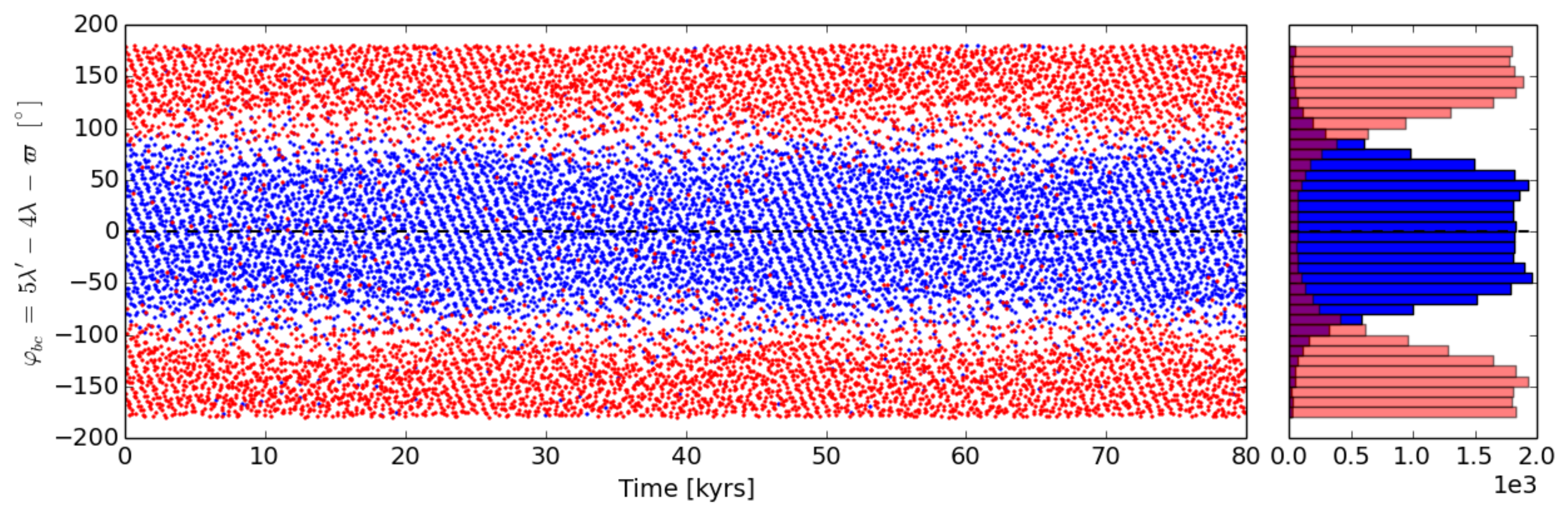}\\
	\includegraphics[width=\textwidth]{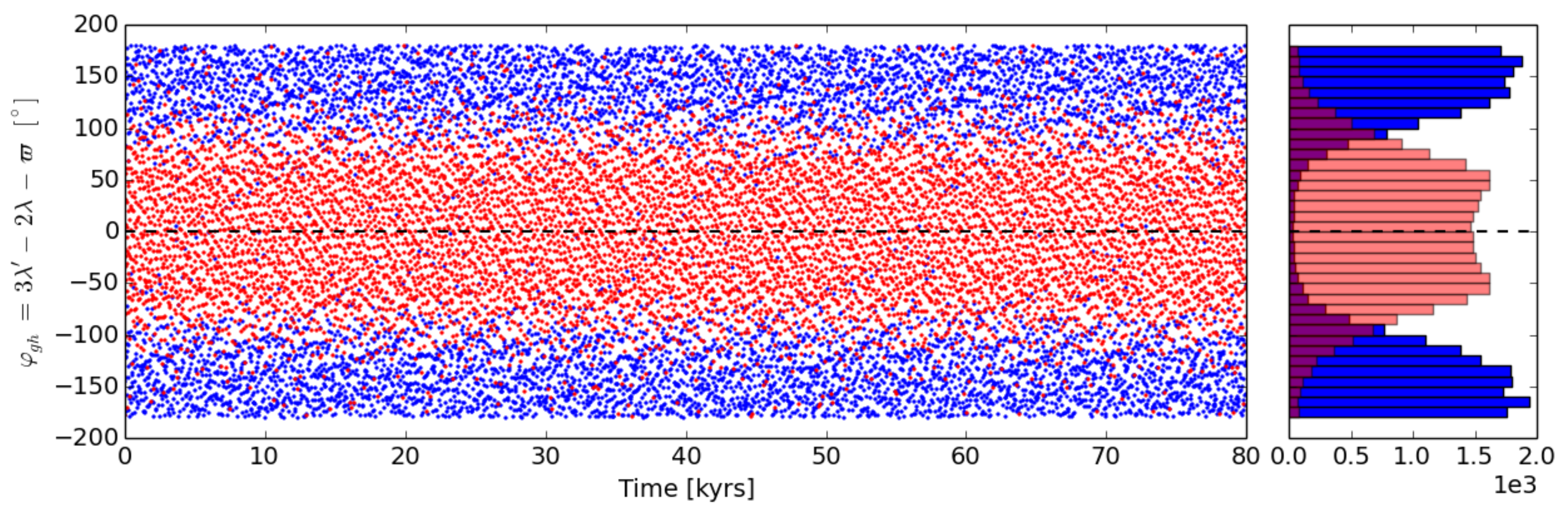}
	\caption{Librating resonant angles in K90. In blue is the interaction with the external planet, $\varphi'$, and in red with the 
			internal, $\varphi$. For clarity, the left panels show the time evolution of the resonant angle only within the first 80 
			kyrs of the simulation; while the right panel (vertical histograms) summarize the behavior of $\varphi'$ and $\varphi$ 
			during 0.3 Myrs. The top panel corresponds to the 5:4 near-MMR between K90b and K90c. The bottom panel shows the 
			resonant angle between K90g and K90h in a 3:2 near-MMR.}
	\label{fig:MMR90}
\end{figure*}

K11 and K90 exhibit TTVs \citep{Lissa11,Cab14}, which occur if there are strong gravitational interactions between the planets. The 
discovery papers of both K11 and K90 discussed MMRs for different planet combinations. The period ratios of K11b-K11c, and K90b-K90c 
suggest a 5:4 near commensurability in both systems. Additionally, \citet{Cab14} discusses a possible 2:3:4 near resonance between 
K90d, K90e and K90f. These first claims of MMR in the system were based on the period ratios of neighboring planets, but dynamically 
a MMR might not occur, due to the dependence on the eccentricity and inclination of the planets. In order to confirm the existence of 
a MMR in a pair of planets it is necessary to estimate their resonant angle. A MMR is present if the resonant angle librates near a 
constant value.

We investigate first-order MMRs between neighboring planets by calculating their resonant angle from the N-body simulations directly 
using
\begin{align}
\varphi' &= j\lambda' + (1-j)\lambda - \varpi' \\
\varphi &= j\lambda' + (1-j)\lambda - \varpi,
\end{align}
where $\lambda$ is the mean longitude and $\varpi$ is the longitude of pericenter. The apostrophe denotes the external planet. Using 
the given masses and orbital elements for K11, all the resonant angle combinations circulate at all times. Thus, we cannot confirm a 
5:4 near-MMR between K11b and K11c. In K90's case, two pairs of neighboring planets display a librating resonant angle (Figure 
\ref{fig:MMR90}). These plots suggest that K90b is in 5:4 near-MMR with K90c and that K90g is in a 3:2 near-MMR with K90h. In both 
cases, the resonant angle librates with large-amplitude around $0^\circ$ and $180^\circ$. The existence of the 5:4 between K90b-c 
suggests that some migration could have taken place. The possible 3-body resonance between d, e and f noted by \citet{Cab14} would 
correspond to a resonant angle $\varphi = 2\lambda_d - 6\lambda_e + 4\lambda_f$. After plotting this angle from our simulations, we 
find that the resonant angle circulates through the length of the realization. The 3:2 near-MMR between K90g and K90h could be the 
origin of the instability in the K90 realizations. The instability in K90 depends on the exact masses of the planets, particularly 
those for K90g and K90h \citep{Cab14}, which are the most massive. Although, the nominal masses of K90 planets are unknown, the impact 
of these massive planets on the system might be strengthened by their apparent MMR.

The behavior of the K11 and K90 analogues explored here show a range of dynamical outcomes that STIPs could have in the presence of 
an outer (undetected) perturber. To date, it is uncertain whether the detected planets among the known exoplanet systems represent a 
complete set or whether there are undetected planets in those systems. Follow up measurements of orbital inclinations of STIPs could 
reveal dynamic rigidity in the orbital plane and as a consequence, could be used as an indirect detection method for an outer planet. 
The observational determination of the longitudes of the nodes of planets in a given STIP could tell us whether there is an external 
perturber should the $\Omega$s for all the planets be $\lesssim 20^\circ$. In a non-perturbed STIP, the nodes would not 
preferentially align to a value, although it might be the case that a subset of the planets present nodal alignment due to coupling. 
Moreover, an external perturber will cause a coherent change in the inclination of the planets, causing the duration of their transits 
to change coherently. An outer planet could also cause large breaks in the nominal orbital plane, which could explain some spin-orbit 
misalignment of low mass planets or reduce the number detected by transit. 

In this paper, we have only considered a single Jovian perturber as a first step. Future work will need to include the effects of multiplicity among outer planetary systems. Having two or more Jovian planets would modify the orbital precession frequencies and increase the complexity of the system' secular frequency spectrum, particularly if the giant planets are mutually interacting. This would have consequences for the dynamics and stability of STIPs and could increase the fraction of inner systems that become unstable. As an example, consider studies based, at least in part, on the solar system, which show that the stability of the inner solar system or inner system analogues can depend sensitively on the orbital configuration of the solar system's outer planets \citep[e.g.,][]{Lithwick11, Agnor12,Clement17}.

In our simulations, the Jovian perturber was placed on a low eccentricity orbit ($e \leq 0.05$). If the perturber's orbital eccentricity were to be increased, which affects the width and power of secular resonances, we would expect additional instability. For example, \citet{Clement17} showed that increasing the nominal eccentricity of Jupiter and Saturn by a factor of two enhances chaos within the inner solar system and reduces the system's stability.

Finally, we integrated our simulations for 10 Myr, and used a corresponding ``migration'' timescale for the perturber. This timescale is a compromise between integrating these systems with sufficiently small time-steps and evolving the systems for a dynamically meaningful duration. In either set of simulations (with or without a perturber), we expect the fraction of unstable systems to increase with time \citep{Volk15}. While we find that the presence of an outer planet does not typically affect the stability of the planetary systems studied here, the migration timescale used to sample different system configurations could have caused the perturber to move through an unstable configuration too quickly, not allowing the instability to develop. As such, the fractional difference between the number of systems that do become unstable with and without an outer planet could increase for longer simulations, and should be explored in future work.\\

\subsection{Comparison with previous and ongoing studies}
The dynamical rigidity of planetary systems has also been observed in the N-body simulations of 55 Cancri \citep{Kaib11,boue14b} and 
HD 20794 \citep{boue14b}. \citet{boue14b} further studied in detail the theory behind the dynamical rigidity of planetary systems in a 
hierarchical setting, in which the host star is part of a binary or an outer giant planet is present, and they explored the conditions 
needed to drive spin-orbit misalignments \citep{boue14a}. 

\citet{Hansen17} also found dynamical rigidity while working on the hypothesis that the \textit{Kepler} Single 
Tranet\footnote{Transiting planet \citep{Tremaine12}.} Excess (KSTE)\footnote{Also known as the Kepler dichotomy.} could be explained 
by secular resonances driven by long-period giant planets. The KSTE is the fractional surplus of single transiting planets based on 
the expected fraction determined from the systems with multiple transiting planets. \citet{Hansen17} found that a fraction of the 
excess could be explained by a mixed population of Jovian and Saturn analogues but at high inclinations. They used in their numerical 
simulations a selection of prototype planetary systems from \citet{Hansen13}, which included systems with multiplicity of 3 to 10 
planets, masses from 1-10 $M_\oplus$ with a mass weighted semi-major axis $<a>_M=0.26-0.5$ au, and an external perturber between 1-5 
au. 

We present a case study of two Kepler systems with high-multiplicity, which is complementary to \citet{Hansen17}. We demonstrate that 
it is possible to drive high inclinations of low-mass planets through secular resonances without a highly inclined perturber and 
confirm that the Lidov-Kozai effect does not occur for STIPs with a massive planetary outer perturber. In our particular case, the 
inclination resonance excited the orbit of the two innermost planets in K90+; a single planet could be driven to a similar outcome 
through secular resonances.\\

\section{Summary}
\label{sec:summary}

In this paper we studied the stability and observability of two known high-multiplicity STIPs, K11 and K90, in the presence of an 
outer Jupiter-like planet through N-body simulations and secular theory. The presence of the perturber causes dynamical rigidity about 
a common orbital plane for the inner planets, while the stability of the system remains unaltered when compared with unperturbed 
realizations for most perturber locations. The observed instability seems to be inherent to STIPs, suggesting secular resonances 
among the planets. The rigid behavior of the orbital plane occurred for most of the parameter space that we explored as long as no 
instability developed. The presence of the perturber also caused two possible effects on systems that are otherwise stable: (1) the 
orbital plane of the planets could be separated into two distinct planes, as in K90+, and (2) the system could become unstable for 
particular perturber locations. The N-body simulations and secular analysis demonstrate that the instability and multiple orbital 
planes are consequences of the eccentricity and inclination secular resonances, respectively. For K11, we suggest that the 
eccentricity resonance close to K11b is the source of the system's inherent instability.

Comparing STIP secular eigenfrequencies to the synthetic counterparts provide a deeper insight into the coupling and possible presence 
of mean motion resonances between planets. K11's nodal precession frequencies indicate dynamical rigidity, seemingly due to K11g, 
although this is dependent on the actual mass of K11g. In K90, our simulations show a 5:4 (near) MMR between K90b and K90c, as well as 
a 3:2 (near) MMR between K90g and K90h. 

Observations of the rigid behavior of a STIP would indicate the existence of an outer planetary system. It is possible that some of 
the detected planetary systems with low multiplicities are part of higher multiplicity STIPs that are affected by secular resonances.

\acknowledgments
We thank B.~Gladman, C.~van Laerhoven, and the anonymous referee for comments that improved this manuscript. This work was supported by an 
NSERC Discovery grant, The University of British Columbia, the Canadian Foundation for Innovation, and the BC Knowledge Development 
Fund. The simulations presented here were enabled in part by support provided by WestGrid (\url{www.westgrid.ca}) and Compute Canada Calcul Canada (\url{www.computecanada.ca}). This research made use of the NASA Exoplanet Archive, which is operated by the California Institute of 
Technology, under contract with the National Aeronautics and Space Administration under the Exoplanet Exploration Program.



\bibliographystyle{aasjournal}
\bibliography{K11-inc}

\end{document}